\documentclass[12pt]{article}
\usepackage{fullpage}
\usepackage{amsmath,amssymb}
\usepackage[dvips]{graphicx}
\usepackage{subfigure}
\usepackage{latexsym}
\usepackage{xcolor,graphicx}
\usepackage{slashed}

\renewcommand{\d}{\text{d}}
\newcommand{\lsuper}[1]{ {}^{#1}\hspace{-2 pt}}

\newcommand{\half}{\frac{1}{2}}
\newcommand{\beq}{\begin{equation}}
\newcommand{\enq}{\end{equation}}

\begin{document}

\begin{center}
{\large \bf  Quantum mechanics as a solution to\\ the  classical self-force problem  }\\

\medskip
Yehonatan Knoll \\

e-mail: \verb"yonatan2806@gmail.com"\\

\begin{abstract}
\noindent It is argued that, contrary to conventional wisdom, no trustworthy  universal self-force/radiative corrections to the Lorentz force equation, can be derived from the basic tenets of classical electrodynamics. This concords with the apparent randomness observed in quantum mechanical scattering experiments and with the absence of any experimental support for such  universality.  In a recent paper \cite{Knoll2017}, the \emph{statistical} effect of radiative corrections to the motion of  charged bodies has been derived from the basic tenets and does take a universal form, described by quantum mechanical wave equations---again conforming with experiment. As that derivation assumes nothing about the size, mass or composition of the body, it is conjectured that quantum mechanics is the appropriate framework for dealing also with radiative corrections to the motion of macroscopic bodies.

\end{abstract}

\end{center}

\section{Introduction}

Classical electrodynamics  of point charges (CE) is neither a valid, an invalid or an approximate theory. It is a non-theory.  The problem is that the electromagnetic (EM) potential is non differentiable exactly where one needs such differentials---on the world lines of particles---and  the electromagnetic  energy density is non integrable there.


The above pathologies notwithstanding, CE   proves to be an immensely practical tool by employing a variety of ad hoc `cheats', applicable in limited domains.     The validity domain of each `cheating method'---and cheating is absolutely necessary for an ill defined mathematical apparatus to produce definite results---is defined \emph{solely by the experimental success of the method}. It is not a sub-domain of the global non-theory in which the latter becomes well defined, nor is it the domain of an approximate theory; CE has no predictions---hence no approximations either---in any domain.

Nevertheless, the results of all those cheating methods can collectively be derived from a concise set of assumptions, dubbed henceforth the \emph{basic tenets of CE}, which are: Maxwell's equations, plus local conservation of a symmetric energy-momentum (e-m) tensor, the EM part of which is the canonical tensor. That is, for all practical purposes, CE is some well defined realization of the basic tenets by means of small charged bodies.

Acknowleging the pivotal role played by the basic tenets,  leads to two approaches to the self-force problem: Explicit and  implicit. In the former,  explicit equations governing the `matter part' of the e-m tensor are postulated (the counterpart of Maxwell's equations, from which the EM part is constructed). We shall review some intuitive such  attempts, showing through their failure  to satisfy the basic tenets, the non triviality of implementing the explicit approach. This, however, does not imply that the task is necessarily all that difficult. One  only needs to  covariantly   define  some phenomenological, short range  attractive force which counters the Coulomb repulsion, thereby allowing for the formation of a stable charged body (nonetheless, the  author knows of no such example, which is clearly much more challenging than in the similar, gravitational case, where gravity naturally supplies the `glue' holding the body together). 

An explicit phenomenological construction, by definition involving  statements about the properties of charged matter, goes beyond what is normally considered to be the domain of CE, and can be challenged in a variety  of  ways. For example, to avoid using the ill defined concept of a point charge, a continuous charge distribution must be used which, in turn, represents a local average of elementary, physical charges. Can a phenomenological construction be considered consistent without proving its compatibility with  the fundamental, explicit construction associated with the constituents of the body? Or perhaps, the latter belongs to the realm of quantum physics? But then, what is the compatibility criterion given the shaky conceptual foundations of QM?   Such questions are usually dismissed as being ``excessively philosophical" in the case of macroscopic bodies but we argue to the contrary. The effect of radiative corrections to Lorentz trajectories, predicated by explicit methods, is so minute for macroscopic bodies, that it has never been directly observed in any experiment. Since countless assumptions are involved in any phenomenological description, both explicit and implicit assumptions, their consistency---mutual and with nature---cannot yet be trusted; It is one of those cases in which the previously mentioned ``cheating methods" is pending experimental confirmation.   And as for microscopic bodies, from electrons to large ions and molecules, the predictions of all explicit models are plain wrong; Local self-force corrections to scattering cross sections, for example, do not reproduce the experimental result predicted by QM. 

It is rather obvious that no mortal can tailor an explicit construction to the huge body of knowledge regarding the nature of matter. If one exists, it must emerge from basic principles. One such construction, dubbed  extended charge dynamics (ECD) \cite{KY}, which is greatly ellaborated upon in the current paper,   is based on the principle of \emph{scale covariance},   a symmetry of CE  which we consider to be just as important as its Poincar{\'e} covariance. Although not (yet) proven to be the unique such construction, it is \emph{extremely} difficult to  merge scale covariance with the basic tenets. It turns out, then,  that in order to merely satisfy the basic tenets in a scale covariant way,  charged (classical) particles must probably be much richer objects than expected, and the universe containing them---a much more bizarre  place. So rich and bizarre, that quantum mechanics (QM) becomes just a natural statistical description of such a classical ontology \cite{Knoll2017}. There are also indications \cite{Knoll_arxiv_2017} that, ECD alone, could be the ontology underlying \emph{all} forms of known matter and that visible matter, thus represented, suffices in explaining the outstanding observations currently requiring for this task the  contrived notions of dark-matter, dark-energy and inflation \cite{Knoll_arxiv_2017}. Most importantly in the context of the self-force problem, ECD conforms with the apparent indeterminism and non locality observed in scattering experiments. Small deviations from classical cross sections (even with big charged molecules), described by QM, are, according to ECD, just small,  radiative corrections to the Lorentz force equation.  It then becomes clear  why these corrections disappear for massive bodies and why they don't for uncharged particles; Those could still have a radiating dipole moment.


The implicit approach to the self-force problem, aims at extracting directly from the basic tenets, irrespective of their realization,  some universal behaviour on the part of accelerating charged bodies. A simple known accomplishment of this approach is that, the Lorentz force equation must be satisfied by \emph{any}, sufficiently massive body, in an external potential which is slowly varying on the scale set by the body's size.  However,   the implicit approach yields  radiative corrections to the Lorentz force equation, having an unspecified  universality domain. The fact that those corrections have never been  demonstrated in an experiment, indicates one of the following: Either additional assumptions entering the analysis, beside the basic tenets, are wrong, or else that the universality domain of radiative corrections has hitherto not been reached in an experiment. In the latter case, their physical relevance should be seriously questioned in light of the huge experimental body of knowledge currently available.

The failure of the implicit approach in describing universal radative corrections to the Loretnz force equation, should be contrasted with the result of \cite{Knoll2017}. Starting with the basic tenets (not necessarily in their ECD realization), a statistical description is derived for an ensemble of bodies, each satisfying them. QM wave equations then emerge as the simplest such description.\footnote{Expressing the ensemble density, $\rho_\text{ens}$, appearing in \cite{Knoll2017} as  $\rho_\text{ens}=\phi^*\phi$, Schr{\"o}dinder's  equation appears as the simplest equation  for $\phi$, consistent with the equations derived $\rho_\text{ens}$.} ECD, then, explains the apparent randomness in QM experiments; The universality of the statistical description appears as a direct consequence of the basic tenets.

The analysis in \cite{Knoll2017} is only approximate, valid for sufficiently small accelerations, but this domain is contained in  the validity domain of all proposed solutions to the self-force problem, whether implicit or explicit.  QED radiative corrections, in this regard, only slightly modify a result which, according to ECD,  already incorporates radiative corrections in an essential way.

\section{Manifestly scale covariant classical electrodynamics}\label{MCE}
The following is a brief review of classical electrodynamics of interacting point charges. It is equivalent to the presentation appearing in any standard book on the matter, but contains a few novel twists.  \\

\noindent Classical electrodynamics of $N$ interacting charges in Minkowski's space ${\textsl M}$ is given by the set of world-lines $\lsuper{k}\gamma_s\equiv \lsuper{k}\gamma(s):{\mathbb R}\mapsto {\textsl M}$, $k=1\ldots N$, parametrized by the Lorentz scalar $s$, and by an EM potential $A$ for which the following action is extremal
\beq\label{I}
I\big[\{\gamma\},A\big]=\int_{\textsl M}\d^4x \left\{\frac{1}{4} F^2+ \sum_{k=1}^N\int_{-\infty}^\infty\d s \,\left( \half \lsuper{k}\dot{\gamma}^2 + qA\cdot \lsuper{k}\dot{\gamma} \right)\delta^{(4)}(x-\lsuper{k}\gamma)\right\}\,.
\enq
Above, $F_{\mu \nu}=\partial_\mu A_\nu -
\partial_\nu A_\mu$  is the antisymmetric Faraday tensor, $q$ some coupling constant, and $F^2\equiv F^{\mu\nu}F_{\mu\nu}$.

Variation of \eqref{I} with respect to any $\gamma$ yields  the Lorentz force equation, governing the motion of a charge in a fixed EM field
\begin{equation}\label{Lorentz_force}
 \ddot{\gamma}^\mu=q\,F^\mu_{\phantom{1} \nu}\dot{\gamma}^\nu\, .
\end{equation} 
Multiplying both sides of \eqref{Lorentz_force} by $\dot{\gamma}_\mu$ and
using the antisymmetry of $F$, we get that $\frac{\d}{\d s}
\dot\gamma^2=0$, hence $\dot\gamma^2$ is conserved by the
$s$-evolution. This is a direct consequence of the
$s$-independence of the Lorentz force, and can also be
expressed as the conservation of a `mass-squared current'
\begin{equation}\label{msc}
b(x)=\int_{-\infty}^{\infty}\d s\,
\delta^{(4)}\left(x-\gamma_s\right)\,\dot{\gamma}_s^2\,
\dot{\gamma}_s\, . 
\end{equation} 
Defining $m= \sqrt{\dot\gamma^2}\equiv \frac{\d
\tau}{\d s}$ with $\tau=\int^s \sqrt{ (\d \gamma)^2 }$ the
proper-time, equation \eqref{Lorentz_force} takes the familiar form
\begin{equation}\label{Lor}
 m\ddot{x}^\mu=q\,F^\mu_{\phantom{1} \nu}\dot{x}^\nu\, ,
\end{equation} with $x(\tau)=\gamma\left(s(\tau)\right)$ above
standing for the same world-line parametrized by proper-time. We
see that the (conserved) effective mass $m$ emerges as a constant
of motion associated with a particular solution rather than
entering the equations as a fixed parameter. Equation
\eqref{Lorentz_force}, however, is more general than \eqref{Lor}, and
supports solutions conserving a negative  $\dot\gamma^2$
(tachyons --- irrespective of their questionable reality) as well as a vanishing  $\dot\gamma^2$.

The second ingredient of classical electrodynamics, obtained by variation of \eqref{I} with respect to $A$, is Maxwell's inhomogeneous equations, prescribing an EM potential given the world-lines of all charges
\begin {equation}\label{Maxwell's_equation}
  \partial_\nu F^{\nu \mu} \equiv \partial^2 A^\mu - \partial^\mu(\partial \cdot A) = \sum_{k=1}^N \lsuper{k}j^\mu \, ,
\end {equation}
with 
\begin{equation}\label{classical_current}
\lsuper{k}j(x)=q\int_{-\infty}^{\infty}\d s\,
\delta^{(4)}\left(x-\lsuper{k}\gamma_s\right)\,
\lsuper{k}\dot{\gamma}_s
\end{equation}
the electric current associated with charge $k$. Applying $\partial_\mu$ to \eqref{Maxwell's_equation} and using the antisymmetry of $F^{\nu\mu}$, implies the conservation of the combined current on the r.h.s. of the equation. This can be explicitly verified for the individual currents $\lsuper{k}j$,
\begin{equation}\label{cons1}
\partial_\mu j^\mu=q\int_{-\infty}^{\infty}\d s\,  \partial_\mu \delta^{(4)}(x-\gamma_s)\dot{\gamma}^\mu_s\, = -q\int_{-\infty}^{\infty}\d s\, \partial_s \delta^{(4)}(x-\gamma_s) = 0\, .
\end{equation}
The current on the r.h.s. of \eqref{Maxwell's_equation} obviously defines $F$ only up to a solution to the homogeneous Maxwell's equation $\partial_\nu F^{\nu \mu}=0$.

\subsection{Scale covariance}\label{ScCo}
The above unorthodox formulation of classical electrodynamics highlights its scale covariance, a much ignored symmetry of CE which, nevertheless, is just as appealing a symmetry  as translational covariance (Poincar{\'e} covariance  in general). Any privileged scale appearing in the description of nature, just like any privileged position, should better be an attribute of a specific solution  and not of the equations themselves which ought to support all properly scaled versions of a solution.  As there seems to be some confusion regarding scale covariance, we try to clarify its exact meaning next.

The Poincar{\'e} group plays a fundamental role in \emph{any} theory, whether covariant or not. In particular, this means that \\
{\bf a.} If some coordinate system is suitable for describing the theory then so is any other system related to the first by a Poincar{\'e} transformation.\\
{\bf b.} Under the above change in  coordinate systems, the parameters of the theory must transform under some representation of the Poincar{\'e} group. Poincar{\'e} covariant theories are those distinguished theories containing only Poincar{\'e} invariant parameters. \\
{\bf c.} The physical content of the theory is identified with invariants of the Poincar{\'e} group, viz., attributes transforming under its trivial representation which are therefore independent of the coordinate system.

Elevating the one-parameter group of scale transformations   to the status of the Poincar{\'e} group amounts  to extending the latter with a  a dilation operation, $x\mapsto \lambda x$ for any $\lambda>0$. By {\bf b} above, we should also assign a \emph{scaling dimension}, $D_\Omega$ to each object, $\Omega$, dictating the latter's transformation under scaling of space-time, $\Omega\mapsto \lambda^{-D_\Omega}\Omega$, and by {\bf c}, only dimensionless quantities have physical meanings (The custom of attaching  `dimensional units' to measurable quantities, such as a kilo or a meter, guarantees that in addition to the scale dependent measurement, another scale dependent gauge is specified, yielding a scale independent ratio). Note, however, that the assignment of scaling dimensions to objects of a theory is not unique unless the theory is scale covariant, viz., contains parameters of scaling dimension zero only (even in this latter case one can distinguish between theories leaving an action invariant  thereby facilitating the derivation of a conserved current associated with scaling symmetry, and those theories only preserving the equations. CE falls into the first category).

Back to the case of classical electrodynamics, we can see that the scaled variables
\begin{equation}\label{st}
 A'(x)= \lambda^{-1}
 A(\lambda^{-1} x)\, , \qquad \gamma'(s)=
\lambda\gamma(\lambda^{-2} s)\,, 
\end{equation}
also solve \eqref{Lorentz_force} and \eqref{Maxwell's_equation}, without scaling of  $q$, hence CE is scale covariant.  From \eqref{st} one can also read the following  scaling dimensions:  $[x]=[\gamma]=1$; $[s]=2$; $[A]=[m]=-1$; $[j]=-3$, and by virtue of scale covariance $[q]=0$.  Poincar{\'e} symmetry combined with  \eqref{st}, forms the symmetry group of CE.

The simplicity in which scale covariance emerges in classical electrodynamics is due to the representation of a charge by a  mathematical point, obviously invariant under scaling of space-time. As we shall see, achieving scale covariance with extended charges is a lot more difficult, as no dimensionful parameter may be introduced into the theory from which the charge can inherit its typical scale.  

\subsection{The basic tenets of CE}\label{CR} 
Associated with each charge is a `matter' energy-momentum (e-m) tensor,  
\beq\label{classical_m}
m^{\nu\mu}=\int_{-\infty}^{\infty}\d s\,\,
 \dot{\gamma}^\nu \, \dot{\gamma}^\mu\,
 \delta^{(4)}\big(x-\gamma_s\big)\, ,
\enq
formally satisfying
\beq\label{ni}
\partial_\nu \lsuper{k}m^{\nu \mu} =F^{\mu\nu}\,\lsuper{k}j_\nu\,,
\enq

\begin{align}
\partial_\nu m^{\nu \mu}&=\int\d s\,\,
 \dot{\gamma}^\nu \, \dot{\gamma}^\mu\,
 \partial_\nu\delta^{(4)}\big(x-\gamma_s\big)=-\int\d s\,\,
  \dot{\gamma}^\mu\,
 \partial_s\delta^{(4)}\big(x-\gamma_s\big)\nonumber\\
 &=\int\d s\,\, \ddot{\gamma}^\mu\,\delta^{(4)}\big(x-\gamma_s\big)=\int\d s\,\,qF^{\mu\nu}\dot{\gamma}_\nu\,\delta^{(4)}\big(x-\gamma_s\big)=F^{\mu\nu}\,j_\nu\,.\nonumber
\end{align}
Likewise, associated with the EM potential is a unique gauge invariant and symmetric\footnote{The symmetry of the e-m tensor is mandatory if it to be a $g_{\mu\nu}\rightarrow\eta_{\mu\nu}$ limit of its general relativistic version as there, symmetry follows from its definition. See \cite{Knoll_arxiv_2017} } EM e-m tensor
\begin {equation}\label{Theta}
\Theta^{\nu \mu}=\frac{1}{4} g^{\nu \mu} F^2 + F^{\nu \rho}
                F_\rho^{\phantom{1} \mu}
\end {equation}  
formally satisfying Poynting's theorem
\begin {equation}\label{Poynting}
 \partial_\nu \Theta^{\nu
 \mu} =- F^{\mu}_{\phantom{1}\nu}\sum_k \lsuper{k}j^\nu   \,,
\end {equation}
where only use of \eqref{Maxwell's_equation} and the identity
\beq \label{id1}
\partial^\mu F^{\nu \rho} + \partial^\nu F^{\rho \mu} +\partial^\rho F^{\mu \nu}=0
\enq
has been made in establishing \eqref{Poynting}. Summing \eqref{ni} over $k$ and adding to \eqref{Poynting} we get a symmetric conserved e-m tensor of the combined matter-radiation system, 
\beq\label{pp}
\,p^{\nu\mu}:=\Theta^{\nu \mu} + \sum_k \lsuper{k}m^{\nu \mu}\quad\Rightarrow\quad \partial_\nu p^{\nu\mu}=0\,, 
\enq
the conservation of which can also be established form the invariance of the action \eqref{I} under translations. Note  that the obvious coupling between matter and radiation notwithstanding, the conserved e-m tensor in \eqref{pp} splits into two pure contributions. Conservation of a generalized angular momentum tensor  follows straightforwardly from the symmetry of $p$ and \eqref{pp}. 

Finally, for future reference, we note that associated with the scaling symmetry \eqref{st} is an interesting conserved `dilatation current' 
\begin{equation}\label{dc}
 \xi^\nu=p^{\nu \mu}x_\mu-\sum_{k=1}^n\,\int\d s\,\delta^{(4)}\left(x-\lsuper{k}\gamma_s\right)s\,\, \lsuper{k}\dot{\gamma}_s^2\,\, \lsuper{k}\dot{\gamma}_s^\nu \, .
\end{equation}
However, the conserved dilatation charge, $\int\d^3\boldsymbol{x}\, \xi^0$, depends on the choice of origin for both space-time, and the $n$ parameterizations of $\lsuper{k}\gamma$, and is therefore difficult to interpret.

Equation \eqref{ni}  and Maxwell's equations \eqref{Maxwell's_equation}, together with electric charge conservation of individual currents $\lsuper{k}j$, are dubbed in this paper the \emph{basic tenets of CE}, and henceforth  shall assume a status of  axioms rather than  derived relations. The rational behind such a step  lies in the fact that, the infinitely detailed dynamics of point charges, or  the singular EM field generated by them, are never the actual subject of observation in experiments to which CE is successfully applied, but rather the basic tenets in their integral forms. For example, the thin tracks left by charges in particle detectors, accurately described by the Lorentz force equation, are consistent with  a hypothetical pair $\{j,m\}$, localized about a common world line, satisfying the basic tenets \eqref{ni} and \eqref{cons1} (see appendix \ref{single}). Likewise, the phenomenon of radiation resistance, whether in wires or particle accelerators,  is a demonstration  of  Poynting's theorem \eqref{Poynting} and e-m conservation \eqref{pp}, and not of a specific, damping self-force resisting the motion of the charges.

As yet another example, consider the EM energy stored in a capacitor or a solenoid. Its subsequent conversion to mechanical energy (heat) in a resistor is a demonstration of \eqref{pp}. Similarly, the  familiar $r_{ab}^{-1}$ dependence of the Coulomb potential between two charged bodies, $a$ and $b$, is hidden in an integral of the cross term ${\boldsymbol E}^{(a)}\cdot{\boldsymbol E}^{(b)}$ part of $\Theta^{00}$ over the entire space (when one assumes that the  self-energy of each charge is a finite constant, and $r_{ab}\gg$ charges' size). Likewise, the potential energy of an infinitesimal circulating current (a magnetic dipole, $d$) in an external field, ${\boldsymbol B}_\text{ext}$, is just due to the ${\boldsymbol B}_\text{ext}\cdot{\boldsymbol B}_d$ term.

The basic tenets are not only verified by \emph{any} experiment---including QM ones---but moreover, it seems impossible for any theory not satisfying the basic tenets to be consistent with the full range of experiments associated even with CE (let alone QM), a small sample of which was described above.


\subsection{The classical self-force problem}\label{self_force}
The self-force problem of CE refers to the fact that the EM potential, $A$, generated by \eqref{Maxwell's_equation} is non differentiable  everywhere on the world  line  $\bar{\gamma}\equiv\cup_s \gamma_s$, traced by  $\gamma$, rendering ill defined the Lorentz force---the r.h.s. of \eqref{Lorentz_force}---as well as the r.h.s. of the constitutive relation \eqref{ni} (even in the distributional sense).  A reminder of this appears in  the form of  non integrable singularities on the $\bar{\gamma}$'s of  the  EM energy density $\Theta^{0 0}$, making the  energy of a system of particles likewise \emph{ill defined}.

Fixing the self-force problem amounts to turning a non-theory into a (mathematically well defined) theory  and there is  no obvious  `right way' of doing so. The simplest way, which often leads to good agreement with experiment, is to eliminate the self generated field from $F$ when computing the Lorenz force acting on a particle. For this to be possible one needs to be able to uniquely define the contribution of each charge to the total field $F$, and the prevailing method is to take the retarded Lienard-Wiechert potential of the charge
\beq\label{kks}
A_\text{ret}(x)=q\int\d s\, \delta\big[\left(x-\gamma_s\right)^2\big]\dot{\gamma}_s\; \theta \left(x^0-\gamma^0_s\right)\,,
\enq      
as that field. The r.h.s. of \eqref{Lorentz_force} is rendered well defined this way, but the basic tenets no longer hold true even in a formal way, their validity follows from the existence of an action, \eqref{I}, not discriminating between the contributions of  different charges to $F$. 


In his celebrated work on the self force problem, \cite{Dirac},  Dirac  attempts to salvage the basic tenets by retaining the self-retarded potential,  writing it as   
\beq\label{ttr}
A_\text{ret}=\half\left( A_\text{ret} + A_\text{adv} \right) + \half\left( A_\text{ret} - A_\text{adv} \right)\,,
\enq
with the advanced Lienard-Wiechert potential
\beq
A_\text{adv}(x)=q\int\d s\, \delta\big[\left(x-\gamma_s\right)^2\big]\dot{\gamma}_s\; \theta \left(x^0+\gamma^0_s\right)\,,\nonumber
\enq    
and ignoring the ill-defined Lorentz force derived from the  first term in \eqref{ttr}. The well defined force derived from the second term modifies the Lorentz force equation into the third order Abraham-Lorentz-Dirac (ALD) equation
\beq\label{ALD}
m\ddot{\gamma}=qF\dot{\gamma}+q^2\frac{2}{3}(\dddot{\gamma} + \ddot{\gamma}^2\dot{\gamma})\,,\quad\dot{()}\equiv\frac{\d}{\d\tau}\,,
\enq
which is Dirac's answer to the self-force problem.
He then tries to justify \eqref{ALD}, basing his argument on the fact that  e-m  is conserved inside an arbitrary tube surrounding any single charge. However, as we show in appendix \ref{single}, Dirac's analysis is flawed; the radius of the tube, ultimately taken to zero by Dirac, cannot serve as a regulator for the singular filed generated by a point-charge. This should not come as a surprise, as the two explicit solutions of \eqref{ALD} he gives in \cite{Dirac}, both formally violate e-m conservation \eqref{pp}. And if one is flexible in defining a point-particle's four-momenta to be other than $m\dot{\gamma}$ \eqref{pp}, then even by Dirac's own admission, e-m conservation alone cannot uniquely define the e.o.m.

In another classic work \cite{Wheeler1}\cite{Wheeler2}, Wheeler and Feynman gave a surprising new look at Dirac's electrodynamics. Elaborating  the formalism of  action-at-a-distance electrodynamics, they found a locally conserved and integrable e-m tensor for a set of  point charges interacting through their half advanced plus half retarded Lienard-Wiechert potentials, without self interaction. Under certain assumptions, a subset of charges surrounded by sufficiently many other charges, behaves in accordance with Dirac's theory. Nevertheless, the form of that integrable EM tensor is  radically different from \eqref{Theta}, admitting both negative values for its energy density component as well as  nonzero values at places where the EM field due to all charges vanishes (implying, among else, gravitational curvature in a generally covariant extension). In fact, the very notion of localization of EM e-m is absent from  Wheeler and Feynman's theory, making it impossible to apply e-m conservation to isolated subsystems---probably the most well tested prediction of CE. Their proposal,  therefore, can hardly be claimed to be consistent with the full range of experiments to which CE is successfully applied. Instead, it is some well defined  theory of interacting point charges sharing with CE a common symmetry group and admitting an integrable and conserved e-m tensor---but it is not CE.

\subsubsection{Extended currents}\label{EC}

Insisting on retaining both the form \eqref{Theta} of the canonical EM tensor and a point charge, inevitably leads to a non-integrable energy density and consequently to violation of the basic tenets. In a second, explicit class of attempts to solve  the self-force problem, one therefore substitutes for the  distributions  \eqref{classical_current} and \eqref{classical_m}  regular  currents, both localized about $\bar{\gamma}$. The regularity of the electric current implies a smooth potential on $\bar{\gamma}$, rendering  the Lorentz force \eqref{Lorentz_force}  well defined and the canonical  EM tensor---integrable.  Various proposals can be found in the literature, all utilizing a `rigid construction' in the sense that the extended currents are uniquely determined by $\gamma$.  This is not only the simplest way to eliminate the singularity of $A$ on $\bar{\gamma}$ but also the only one allowing to retain the Lorentz force equation \eqref{Lorentz_force}. Below, we shall employ a rigid construction which is equivalent to the one employed in \cite{McManus} but via a different method which will serve us later.

The idea is to substitute for $\delta^{(4)}$ in \eqref{classical_current} a finite approximation of a delta function, respecting the symmetries of the theory. In Euclidean four dimensional space this is straightforward: $\delta^{(4)}(x)\mapsto a^{-4}h(x/a)$ for any normalized spherically symmetric $h$ and some small $a$. In Minkowski's space this is more tricky due to the non-compactness of Lorentz invariant manifolds $x^2=const$, so first we note that the current
\beq\label{gkg}
\int\d s\, \frac{1}{\epsilon}h\left[\frac{\left(x-\gamma_s\right)^2}{\epsilon}\right]\dot{\gamma}_s\,,
\enq
is conserved and significantly differs from the $\epsilon$-independent  current
\beq
\int\d s\, \delta\left[\left(x-\gamma_s\right)^2\right]\dot{\gamma}_s\,,
\enq
only up to a  distance from $\gamma_s$ on the order of $\sqrt{\epsilon}$  (in the rest frame of $\gamma_s$). Taking the derivative of \eqref{gkg} with respect to $\epsilon$ we therefore get a conserved current 
\beq\label{pre_ECD_current}
j(x)=q\frac{\partial}{ \partial \epsilon}\int\d s\, \frac{1}{\epsilon}h\left[\frac{\left(x-\gamma_s\right)^2}{\epsilon}\right]\dot{\gamma}_s\,,
\enq
which is significant only inside a ball of radius $\sim\sqrt{\epsilon}$ in the rest frame of $\gamma$, reducing to the line current \eqref{classical_current} in the limit $\epsilon\rightarrow 0$ for a properly normalized $h$. Pushing the derivative into the integral, the regular function
\beq\label{fdhg}
\frac{\partial}{ \partial \epsilon} \frac{1}{\epsilon}h\left(\frac{x^2}{\epsilon}\right)\,,
\enq
appears (up to a normalization constant) as a finite approximation to the invariant $\delta^{(4)}(x)$ entering \eqref{classical_current}. This can indeed be directly verified. Note, however, that even for a compactly supported $h$, \eqref{fdhg} is non vanishing in some neighborhood of the light-cone $x^2=0$ for an arbitrarily large (light like) $x$.\footnote{The `pickup' property of \eqref{fdhg} is achieved by means of  its rapid oscillation across the light cone, i.e., near large light-like $x$, \eqref{fdhg} takes both positive and negative values.} Consequently, the current \eqref{pre_ECD_current} is never compactly supported and can be shown to have an (integrable) algebraically decaying `halo'.  We see that the obvious way of covariantly generalizing Lorentz's construction of a finite-size electron, leads to  weakly localized currents. 

Choosing the retarded solution of Maxwell's equations with the current \eqref{pre_ECD_current} as source, one arrives at the following expansion for the self-force correction to the Lorentz force at $\gamma_s$, when $s$ is chosen as the proper time:
\beq\label{dds}
-\frac{q^2 C}{\sqrt{\epsilon}}\ddot{\gamma} +  f_\text{ALD} + q^2O(\sqrt \epsilon)\,.
\enq 
Above,  $C$ is some positive constant depending only on $h$,   the $O(\sqrt{\epsilon})$ term also depends on the local form of $\gamma$, and $f_\text{ALD}$ stands for the standard Abraham-Lorentz-Dirac  self-force, $q^2\frac{2}{3}(\dddot{\gamma} + \ddot{\gamma}^2\dot{\gamma})$.

There are three major difficulties with the above extended current approach to the self-force problem. First, it introduces an arbitrary function---an infinite set of parameters---into single-parameter CE. Second, the dimensionful parameter $\epsilon$ spoils the scale-covariance of CE. Finally, the basic tenets are still not satisfied, the problem being with  \eqref{ni} whose l.h.s.  is supported  on $\bar{\gamma}$ while its r.h.s. now extends beyond it.
Taking the ($\epsilon\rightarrow 0$) limit of a point-charge distribution results in an effective infinite mass, the first term in \eqref{dds}. The textbook reply to that divergence is choosing an oppositely diverging `bare mass' such that the sum remains positive and finite. This, however, only shifts the original self-force problem from being that of an ill-defined  Lorentz force to an ill-defined (singular) basic tenet \eqref{ni}. 

We can be more careful by choosing a most general, covariant,  non singular ansatz for $m^{\mu\nu}$
\beq\label{pre_ECD_m}
m^{\mu \nu}(x)=\frac{\partial}{ \partial \epsilon}\int \frac{1}{\epsilon}h_1\left[\frac{\left(x-\gamma_\tau\right)^2}{\epsilon}\right]\dot{\gamma}^\mu\dot{\gamma}^{\nu}+g^{\mu\nu} \frac{1}{\epsilon}h_2\left[\frac{\left(x-\gamma_\tau\right)^2}{\epsilon}\right]\dot{\gamma}^2\d \tau\,,
\enq
for some functions $h_1$, $h_2$, chosen such that \eqref{ni} is satisfied for a freely moving particle, viz., in a vanishing external filed (e.g. as  implicitly provides by \cite{Schwinger}).    In the presence of a non vanishing external field,  the value of the l.h.s. of   \eqref{ni} at any $x$ depends only on the value of the external field  on $\bar{\gamma}$, whereas the r.h.s. depends also on its  local value at $x$. No choice of $h$'s  can lead to point-wise equality in \eqref{ni}, so we still need to take  the $\epsilon\rightarrow 0$ limit,  restricting the support of both sides of \eqref{ni} to $\bar{\gamma}$ which, in turn, implies  negative, $O(1/\sqrt{\epsilon})$ divergent $h$'s for a finite mass. However, while taking smaller a $\epsilon$ restricts \eqref{ni}-violation possible only in an $\sqrt{\epsilon}$-neighborhood  $\bar{\gamma}$, the degree to which this happens is expected to explode due to the Coulomb divergence and that of \eqref{pre_ECD_m}, hence the distributions substituting both sides of \eqref{ni} in the limit are not expected to be identical.     Verifying this requires `zooming' into the singularity which develops on $\bar{\gamma}$ in the limit, which is not a trivial task. Instead, towards the end of appendix \ref{single} we demonstrate explicitly why the ALD equation,  applied to a group of interacting charges, violates \emph{global} e-m conservation. 


Yet another failed attempt to reconcile  the ALD equation with e-m conservation, involves rescaling the charge and (bare) mass, $q\mapsto\sqrt{\epsilon}\,q$, $m\mapsto\sqrt{\epsilon}\,m$. Adding \eqref{dds} with the scaled charge to a Lorenz force equation with scaled parameters,  and dividing by $\sqrt{\epsilon}$, the zero order effect is  a finite renormalization of the mass, $m\mapsto m+q^2C$, in the original  Lorenz force equation. The `universality' of the ALD self-force in this case, is equivalent to the statement that $\sqrt{\epsilon}f_\text{ALD}$ is the leading correction to the Lorentz force in an $\epsilon$ expansion (note the $\sqrt{\epsilon}$ multiplying the standard ALD force!). The  $\epsilon\rightarrow 0$ limit is  void of any  physical meaning, as  both the charge and the mass, though retaining a fixed ratio in the limit, vanish. Consequently, the dynamics of a group of such interacting charges must trivialize to uniform motion---as in our original $\epsilon\rightarrow 0$ limit.  An implicit approach, leading to the same result  appears in \cite{Wald}. However, the counterpart of the  $O(\sqrt{\epsilon})$ term, in this case, cannot be computed explicitly, and this is no surprise; As we further show towards the end of appendix \ref{single}, no local correction to the Lorentz force equation can guarantee e-m conservation when only retarded fields are included.

If a physically realistic  $\epsilon$ is used instead of the limit, and the effects of the ALD term (e.g., via the reduced order ALD equation; see \cite{Wald}) are still undetectable in an experiment, this could only mean that, either the assumption of using only retarded solutions (which is at odd with ECD) is inconsistent with nature's realization of the basic tenets, or that higher order corrections are not negligible, meaning that whatever extra assumptions were used to bound the higher order corrections, are invalid.   The fact that  no experiment has ever found such universal deviations from Lorentz trajectories,  renders the physical relevance of the implicit approach moot.

Summarizing, CE of point charges cannot satisfy the basic tenets, if only because of the singularity in the canonical tensor,   while  CE of rigid extended charges, though eliminating the former problem,  further spoils scale covariance and introduces infinitely many new parameters. The implicit approach, while  rigorous, has no specific validity domain and is therefore experimentally moot.    To these approaches to the self-force problem one may add nonlinear electrodynamics, notably the Born-Infeld version, in which the singularity in a (modified) canonical EM tensor is rendered integrable at the cost of modifying Maxwell's equations and making them nonlinear. The potential $A$ is  non-differentiable at the source hence the paths of charges are still ill-defined in those theories which further suffer from broken scale covariance, and manifestly violate the basic tenets in their experimentally established form. Finally, solitary solutions of non linear PDE's, as possible solutions to the self-force problem, are ruled out below. 

\section{Extended Charge Dynamics}\label{Extended current}

Our starting point in the construction of currents satisfying the basic tenets  is the electric current \eqref{pre_ECD_current} and expression \eqref{pre_ECD_m} for the e-m tensor $m$. We saw above that the `rigidity' of the covariant integrands in both currents leads to violation of the basic tenets, while their nonsingular nature further spoils scale covariance. To fix both problems we substitute for them  more `vibrant' integrands  which \emph{do} depend on the local field $F$, and whose characteristic scale surfaces naturally without introducing extra dimensionfull parameters. To this end, let us look at the  proper-time Schr\"{o}dinger
equation (also
known as a five dimensional Schr\"{o}dinger equation, or Stueckelberg's equation),
\begin {equation}\label{Schrodinger}
 \Big[ i\bar{h} \partial_s - {\cal H}(x)
 \Big] \phi(x,s)= 0\,, \qquad\text{ }\; {\cal H}=-\frac{1}{2}D^2+V\, ,
\end {equation}
with
\begin {equation}\label{covariant derivative}
 D_\mu = \bar{h}\partial_\mu - i q A_\mu
\end {equation}
the gauge covariant derivative, $A$ and $V$ some vector and scalar potentials respectively, $\bar{h}$ a real dimensionless `quantum parameter', not to be confused with $\hbar$, and $q$ some EM coupling constant. It can be shown by standard means that solutions of \eqref{Schrodinger} satisfy a continuity equation
\beq\label{gsm}
\partial_s \rho=\partial\cdot J\,,\quad\text{with } J=q\,\text{Im }\phi^* D
 \, \phi\,,\quad \rho=q\left|\phi\right|^2\,,
\enq
and  four relations 
\begin{align}\label{ksh}
&q^{-1}\partial_s J^\mu = F^{\mu \nu}J_\nu +q^{-1}\partial^\mu V\rho - \partial_\nu M^{\nu \mu}\,,\\
&\text{with}\quad M^{\nu\mu}=g^{\nu \mu} \left( \frac{i \bar{h}}{2}\left(  \phi^*\partial_s \phi\,
 - \partial_s \phi^* \phi \right)   -
  \frac{1}{2}\left(D^\lambda \phi\right)^* D_\lambda \phi \right) + \frac{1}{2}\big(D^\nu
               \phi \left(D^\mu \phi\right)^* + \text{c.c.}
               \big)\,.\nonumber 
\end{align} 
The common implications of the  non relativistic counterparts of \eqref{gsm} and \eqref{ksh} are probability conservation and Ehrenfest's theorem, and readily carry to the relativistic case. Multiplying \eqref{gsm} by $q^{-1}x$ and integrating its r.h.s. by parts over four-space, we get (for a normalized $\phi$, but this is immaterial to the result we wish to establish) the four momentum of a wave packet. Integrating \eqref{ksh} over four-space, and substituting the above momentum for the l.h.s., localized wave-packets can  then be shown to  trace classical paths when the EM field varies slowly over their extent and the scalar potential $V$ vanishes.

Yet, another implication of  \eqref{gsm} and \eqref{ksh} which has no direct nonrelativistic counterpart is obtained by integrating the two equations over $s$ rather than space-time. The $s$-independent current
\beq\label{jJ}
 j(x)=\int_{-\infty}^{\infty}\d s\, J(x,s)
\enq 
is conserved and, for a vanishing $V$, the constitutive relation \eqref{ni} is satisfied by $j$ and
\beq\label{mM}
 m(x)=\int_{-\infty}^{\infty}\d s\, M(x,s).
\enq 
Associating a unique $\phi$ with each particle and taking the sum of the corresponding currents, $j$,  as the source of Maxwell's equations \eqref{Maxwell's_equation}, the basic tenets are fully satisfied, and the full symmetry group---scale covariance in particular---is retained. 

The above realization of the basic tenets, nevertheless, is apparently inconsistent with the condition of localized $j$ and $m$. The dispersion inherent in the  Schr{\"o}dinger evolution \eqref{Schrodinger} implies that a localized wave-packet gradually spreads even in a potential free space-time. In collisions with an external potential the situation is even worse, and may result in a rapid loss of localization. This means that the  wave-packet could maintain its localization under the $s$-evolution \eqref{Schrodinger} only if somehow the EM potential generated by its associated current $j$, creates a binding trap, but the prospects of such a solution are dim as the self generated Coulomb potential is repulsive rather than attractive. It is further unlikely that such a  self-trapping solution, even if it exists in some otherwise potential free region of space-time, would retain its localization following violent (realistic) interactions with EM potential generated by other charges. Finally, it can be shown  that equation \eqref{Schrodinger} and its associated currents admit a much more natural interpretation in terms of an ensemble of particles, making the single particle interpretation seem rather contrived.

It appears inevitable that  for \eqref{Schrodinger} to be useful in the realization of the basic tenets by means of localized currents, an additional localization mechanism for the wave packet must be introduced into the formalism. In  \cite{KY}, this mechanism takes the form of  a (point) `delta function potential', $V=\delta^{(4)}(x-\gamma_s)$, moving along some $\bar{\gamma}$ in Minkowski's space, which is plugged into the Hamiltonian in \eqref{Schrodinger}, preventing   the wave function  from spreading by the binding action of the potential. Note that essentially any other choice of  binding potential would lead to violation of scale covariance.

\subsection{The central ECD system}\label{central ECD system}
In order for the previous results established for the case $V\equiv0$, viz., Ehrenfest's theorem and the basic tenets, to still hold true for $V=\delta^{(4)}(x-\gamma_s)$, we need to have  $\partial\delta^{(4)}(x-\gamma_s)\rho=0$. As both  Ehrenfest's theorem and the basic tenets in their integral form (the only form directly related to physical observables) involve an integral over four-space, our proviso is equivalent to $\partial\rho\left(\gamma_s, s\right)=0$. Associated with each particle, then, is a pair $\{\lsuper{k}\phi,\lsuper{k}\gamma\}$, $k=1\ldots N$, performing a tightly coordinated `dance': $\gamma$ points to   $\phi$ where to focus, but simultaneously follows the extremum of its muduls squared.  In mathematical terms this dance takes the form of  two coupled equations dubbed the \emph{central ECD system}. The first is an integral version of \eqref{Schrodinger} containing a delta function potential   (see \cite{KY} for a formal derivation) and reads (omitting the particle index on $\phi$ and $\gamma$) 
\begin{align}\label{integral_delta_equation}
 \phi(x,s)&=-2\pi^2\bar{h}^2\epsilon i  \int_{-\infty}^{s-\epsilon} \text{d}s'\,
  \,G(x,\gamma_{s'};s-s') \phi(\gamma_{s'},s')\\\nonumber
  &+2\pi^2\bar{h}^2\epsilon i \int_{s+\epsilon}^{\infty} \text{d}s'\,
  \,G(x,\gamma_{s'};s-s') \phi(\gamma_{s'},s')\\ \nonumber
  &\equiv -2\pi^2\bar{h}^2\epsilon i  \int_{-\infty}^{\infty}\text{d}s'\,
  \,G(x,\gamma_{s'};s-s') \phi(\gamma_{s'},s'){\mathcal
  U}(\epsilon;s-s')\, ,
\end{align}
\begin {equation} \text{with}\qquad {\mathcal
U}(\epsilon;\sigma) = \theta(\sigma-\epsilon) -\theta(-\sigma -
\epsilon)\nonumber\, , 
\end {equation} 
and the second equation is naturally
\begin {equation}\label{surfing_equation}
\partial_x\left|\phi(x,s)\right|^2
 \big|_{x=\gamma_s}\equiv\partial_x\left|\phi(\gamma_s,s)\right|^2=0\, .
\end {equation}

Above, $G(x,x';s)$  is the \emph{propagator}
of  a   proper-time Schr\"{o}dinger
equation, viz., solution of \eqref{Schrodinger} 
satisfying the initial condition (in the distributional sense),
\begin {equation}\label{init_condition}
 G(x,x';s)\underset{s \rightarrow 0}{\longrightarrow}\delta^{(4)}(x-x')\, .
\end {equation}
The extra parameter, $\epsilon$, of dimension $2$, which is needed for the construction of the scale-invariant delta function potential, is ultimately taken to zero and is discussed below.

A delta function potential, inserted into a differential equation, cannot possibly go as smoothly as presented hitherto. Indeed, the central ECD system, first derived in \cite{KY}, is just a background intuition for turning that formal potential into a well defined mathematical object and it turns out that the best way to do so is to look at the basic tenets. As $\phi(x,s)$ solving  Schr{\"o}dinger's equation \eqref{Schrodinger} in the presence of a point potential $\delta^{(4)}(x-\gamma_s)$,  is formally a solution of the  free Schr{\"o}dinger equation \eqref{Schrodinger} for any $(x,s)\neq (\gamma_s,s)$, it appears that the basic tenets would be respected by $j$ \eqref{jJ}, and $m$ \eqref{mM}, for an \emph{arbitrary} $\gamma$ and $x\notin \bar{\gamma}$. The only way, therefore, for the basic tenets to be sensitive to the choice of $\gamma$ (the path taken by the `center' of the particle!) is if the exclusion of $\bar{\gamma}$ from their domain somehow affects their validity. And indeed, as shown in the appendix,  a `wrong' choice of $\gamma$ leads to `leakage' of mechanical e-m associated with $m$, and to electric charge associated with $j$, to  `world-sinks' on $\bar{\gamma}$, rendering the (local) basic tenets useless by preventing their conversion  into integral conservation laws via Stoke's theorem. The derivation of the `fine tuned' central ECD system, shown in the appendix, is guided by this no-leakage criteria.  

Returning to the extra parameter, $\epsilon$,  appearing in \eqref{integral_delta_equation}, in the appendix we show that the $\epsilon\rightarrow 0$ limit of a family of solutions to the $\epsilon$-dependent central ECD system, indeed exists, but not without the usual toll paid for manipulating formal mathematical objects. For it turns out, that solutions of the central ECD system  develop a distribution on the light cone of $\gamma_s$ in the limit $\epsilon\rightarrow 0$. As both $J$ and $M$---the integrands of  $j$ \eqref{jJ}, and $m$ \eqref{mM}, respectively---are bilinears in $\phi$ and its adjoint, a meaningless product of two distributions is formed as a result of taking the $\epsilon \rightarrow 0$ limit of $\phi$ and only then plugging it into  $J$ and $M$. A similar product of distributions is the source of much of the troubles in QFT and is overcome by  two steps: covariant regularization of the distributions,  followed by `renormalization', viz., making sense of possible infinities arising from the removal of the regulator. Likewise, in ECD a covariant regulator, $\epsilon$, is built into the formalism, and the counterpart of the renormalization step    takes the form of a  simple covariant prescription
\begin {equation}\label{ECD_current}
 j\mapsto\lim_{\epsilon\rightarrow 0}\frac{\partial}{\partial \epsilon}\epsilon^{-1}j\, ,\qquad x\notin \bar{\gamma}
\end {equation}
\begin {align}\label{ECD_m}
   m\mapsto\lim_{\epsilon\rightarrow 0}\frac{\partial}{\partial \epsilon}\epsilon^{-1} m\,,\qquad x\notin \bar{\gamma} \,,
\end {align} 
namely, plug a finite-$\epsilon$ $\phi$ into $j$ and $m$, apply a certain `infinity removal' operation: $\partial_\epsilon\epsilon^{-1}$ (see appendix \ref{refined_ECD} for details), taking the $\epsilon\rightarrow 0 $ limit only as a final step. This trick yields a locally conserved smooth electric current which does not leak to a world-sink on $\bar{\gamma}$ and is integrable. Likewise, the e-m tensor of a system of interacting particles, $\Theta + \sum_k \lsuper{k} m$, is locally conserved, non leaking and integrable. It follows that Stoke's theorem can be freely applied to the local basic tenets, as if their domain included all of space-time.

{\bf Spin.} A lot has been said about non-integer spin being one of the hallmarks of QM, but while the above procedure of realizing the basic tenets involves  scalar $\lsuper{k}\phi$'s, a similar method  exists in which  each  $\lsuper{k}\phi$ transforms under an arbitrary representation  of the Lorentz group.  The spin of a particle is a nonphysical  `label' of  the particular method used to construct  $j$ and $m$, both transforming under integer representations of the Lorentz group, whose internal such currents depend on  that method.    An example of spin-$\half$ ECD is discussed in appendix \ref{Spin-half}.

\subsection{The nature of particles in ECD}\label{nature}
The simplest possible problem in ECD is that of single a stationary particle in an otherwise void universe. That is, \emph{the very existence of a particle is due to a  nontrivial localized solution}, viz. $A\neq0$ up to a gauge transformation, for the coupled ECD-Maxwell system. Using a small-$\bar{h}$ approximation  of the propagator, we show in appendix \ref{The classical limit} that such solutions must indeed be particle-like, represented by integrable currents which are   localized about their center $\bar{\gamma}$, and this conclusion is not an artifact of the small-$\bar{h}$ analysis but rather a direct consequence of equation \eqref{integral_delta_equation}.

In a naive approach, finding a particle solution in ECD amounts to guessing a potential $A$, then solving the central ECD system \eqref{integral_delta_equation},\eqref{surfing_equation} for a pair, $\{\phi,\gamma\}$, from which the electric current \eqref{ECD_current} is computed, and `hoping' that this current, along with the initial guess, $A$, indeed solves Maxwell's equation \eqref{Maxwell's_equation}.

By the scale covariance of ECD, to each such isolated solution there corresponds an infinite family of scaled versions, sharing the same electric charge and spin but differing on their self energy which has dimension $-1$. It is  hypothesized that different elementary particles are just scaled versions of each other, hence their common charge. A possible explanation for the observed  `spontaneous scaling symmetry breaking', viz., the absence of an observed continuum of masses, and its dramatic implications to cosmology, appear in \cite{Knoll_arxiv_2017}.


An elementary particle solution, or any other solution for that matter, must come with an `antiparticle' solution to the ECD equations. This is a consequence of  the symmetry of ECD under a `CPT'  transformation:
\begin{align}\label{CPT}
&A(x)\mapsto -A(-x)\,,\quad \gamma(s)\mapsto -\gamma(s)\,\quad \phi(x,s)\mapsto\phi(-x,s)\nonumber\\ 
&\quad \Rightarrow\;j(x)\mapsto -j(-x)\,,\quad m(x)\mapsto m(-x)\,.
\end{align} 

\subsubsection{Comparison with solitons}
The possibility of representing elementary particles by solitary solutions of nonlinear PDE's has been extensively studied in the past. Coupled Maxwell--Dirac \cite{Radford2} or Maxwell--Klein-Gordon \cite{GEORGIEV2005957} systems (some also add a nonlinearity to either the KG or Dirac equations, whose purpose is not entirely clear as the original system is highly nonlinear to begin with) can even be shown to posses spherically symmetric localized solutions which satisfy the basic tenets, although the mass term spoils scale-covariance. Classically speaking,  a self-trapping `charged dust cloud' is obviously impossible due to the repulsive Coulomb self-force (one cannot add a non-electromagnetic force  countering this repulsion without  violating the basic tenets and further breaching scale-covariance) and it turns out that this intuition also applies to such  coupled systems. To counter the Coulomb self-repulsion,  one must therefore add  a point-charge with an opposite sign at the center of the soliton  (or, equivalently,  impose boundary conditions at the origin, forcing the radial electric field there to diverges as $r^{-2}$). This means that the electric field at the origin behaves as baldly as in the point charge case, rendering the self-energy ill-defined. Moreover, the monopole of that solution can be shown to vanish so charged particles cannot be represented by such solutions anyhow (see \cite{Radford2} and references therein). A similir singularity is also found in \cite{Temnenko}, where the  ansatz $m^{\mu \nu}=aj^\mu j^\nu +b g^{\mu\nu}j^2$ is plugged into \eqref{ni}\eqref{Maxwell's_equation} and solved for a spherically symmetric $j$.  It should be noted in this regard that ECD charged particles likewise have a divergent electric field on $\bar{\gamma}$, but this field leads to an integrable self-energy. Moreover, the dynamics of ECD particles is based on the condition that no electric charge nor e-m leakage occurs at those singularities---an analysis which is missing altogether from \cite{Radford2} and related work. 

An apparent way out of this dead end is to add gravity, leading to a coupled Einstein--Maxwell--Dirac system, satisfying the generally covariant basic tenets. Given the relative weakness of gravitational attraction  compared with electrostatic repulsion, it is rather surprising that such localized, non-singular solutions actually exist \cite{Finster1999431}. Nevertheless, the stability of those (static) solutions is only demonstrated for a limited class of---small, by definition---perturbations. More generally, the very ambition  of modelling particles, which are objects maintaining their localization and identity notwithstanding \emph{strong} interaction with the environment, by means of fields, extending throughout space-time, without introducing a focusing/stabilising center, such as $\gamma$ in ECD, seems like an entirely hopeless program---a point we have already made with regard to  possible Maxwell--Stueckelberg solitons, discussed at the end of section \ref{Extended current}.

\subsection{The necessity for advanced solutions of Maxwell's equations}\label{FDM}
In a universe in which no particles imply no EM field, a solution of   Maxwell's equations is  uniquely determined by the conserved current, $j$, due to all particles. The most general such dependence which is both Lorentz and gauge covariant takes the form  
\begin {equation}\label{convolution with K}
 A^\mu(x)=\int d^4 x' \big[\alpha_\text{ret}(x') K_\text{ret}{}^{\mu \nu}(x-x') + \alpha_\text{adv}(x')  K_\text{adv}{}^{\mu \nu}(x-x')\big]j_\nu(x')\,, 
\end {equation}
for some (Lorentz invariant) space-time dependent functionals, $\alpha$'s, of the current $j$, constrained by  $\alpha_\text{ret}+\alpha_\text{adv}\equiv1$, where  $K_{{}^\text{ret}_\text{adv}}$ are the advanced and retarded   Green's function of \eqref{Maxwell's_equation}, defined by \footnote{More accurately, \eqref{K defined} and \eqref{causality condition} do not uniquely define $K$ but the remaining freedom can be shown to translate via \eqref{convolution with K} to  a gauge transformation $A\mapsto A +\partial \Lambda$, consistent with the  gauge covariance of  ECD.}
\begin {equation}\label{K defined}
\left(g_{\mu \nu} \partial^2 - \partial _\mu \partial_\nu\right) K_{{}^\text{ret}_\text{adv}}{}^{\nu
\lambda}(x)=g_\mu^{\phantom{1}\lambda} \delta^{(4)}(x)\,,
\end {equation}
\begin {equation}\label{causality condition}
 K_{{}^\text{ret}_\text{adv}}(x)=0\,\,\,\ \text{for }\, x^0 \lessgtr 0\,.
\end {equation}
In ill defined CE of section \ref{MCE}, $\alpha_\text{adv}\equiv 0$ is taken as a \emph{definition}.  Modulo the self force problem, the fact that CE admits a formulation in terms of  a Cauchy initial value problem (IVP) means that indeed, solutions of CE may be found containing only retarded fields. This proviso, however, is incompatible with the ECD equations, and  $\alpha_\text{adv}$   would generally differ from zero and  vary across space-time.   In particular, the fact that the ECD current  also depends on $A$, both explicitly through the gauge covariant derivative $D$, and implicitly via $\phi$'s dependence on $A$, means that the solution of even a single radiating ECD particle must include  advanced components as these cannot be eliminated by the addition of a solution of the homogeneous Maxwell's equations, as in CE. More generally, the values of both $\alpha$'s must be \emph{read} from a global ECD solution, involving both fields and currents, rather than being \emph{imposed} on it.

That advanced solutions of Maxwell's equations are on equal footing with retarded ones is outraging from the perspective of the (almost) consensual paradigm which accepts only retarded solutions as physically meaningful. One can think of two major reasons for this outrage. The first is the  parallelism which which is often drawn with `contrived'  advanced solutions of other physical wave equations (e.g. surface waves in a pond converging on a point and ejecting a pebble). This parallelism, however, is a blatant  repetition of the historical  mistake which led to the  invention of the aether. The formal mathematical similarity between the d'Alembertian---the only linear, Lorentz invariant second-order differential operator---and other (suitably scaled) wave operators, is no more than a misfortunate coincidence. Has this coincidence  had some real substance to it, then application of the  Lorentz transformation to the wave equation describing the propagation of sound, for example, would have yielded a meaningful result. It is quit remarkable that over a century after the existence of the aether was refuted, and the geometrization revolution of Minkowski in mind, terms such as `wave' and 'propagation' are still as widely used in the context of electromagnetic phenomena as in the  nineteenth century. 

The second, stronger case for rejecting advanced solutions is observational. While at the microscopic scale we challenge this assertion in \cite{Knoll2017} and \cite{Knoll_arxiv_2017}, indeed, no macroscopic object is observed anywhere  spontaneously increasing its energy content by the convergence of advanced radiation on it. Nevertheless,  this stronger argument does not imply that no advanced fields are involved even in such  macroscopic processes,  but rather that our universe has a  \emph{macroscopic radiation arrow-of-time}, discussed next. When, for example, a LED converts the electrostatic energy, previously  stored in a capacitor, into light which, in turn, recharges a second capacitor via a (perfectly efficient...) photoelectric cell, we can only deduce what is the imbalance between the advanced and retarded Poynting fluxes integrated across a surface containing either systems  (and that the two integrals are equal in magnitude but with opposite signs).  

There remains the question of why there is a macroscopic  arrow-of-time, and why the observed direction rather than the reversed. One legitimate answer relies on the anthropic principle: Without the observed arrow-of-time (its direction included) which could very well be just a peculiarity of a  specific ECD solution, we wouldn't exist to raise the question. Alternatively, it may turn out that when global cosmological considerations are included in the analysis, only the observed arrow-of-time becomes possible.
But there is another possibility, more specific to ECD: The coupled ECD-Maxwell system is \emph{not} covariant with respect to time-reversal; CPT symmetry \eqref{CPT} is the closest one gets to the notion of `running the movie backward'. The observed direction of the arrow-of-time is therefore intimately linked with the manifest imbalance between particles and their antiparticles. \\

\noindent {\bf Acknowledgements.} Many thanks to Ofek Birnholtz and Amos Ori for  useful discussions on the self-force problem.

\begin{center}
{\Large \bf Appendices}
\end{center}
\appendix

\section{The `fine tuned' central ECD system}\label{refined_ECD} 

As all ECD currents are computed in the limit $\epsilon \rightarrow 0$, the central ECD system  \eqref{integral_delta_equation} and \eqref{surfing_equation} is given bellow  an operational definition for small $\epsilon$ only. To this end, we would need the small-$s$ form of the propagator $G$. 
Plugging the ansatz
\beq\label{short_s_propagator}
G(x,x',s)=G_\text{f}e^{i \Phi(x,x',s)/\bar{h}}
\enq
into \eqref{Schrodinger}, with 
\beq
G_\text{f}(x,x';s)=\frac{i }{(2\pi\bar{h})^2}\frac{ e^{ \frac{
i(x-x')^2}{2\bar{h}s}}}{s^2}\, \text{sign} (s)\,,
\enq
the free propagator computed for $A\equiv 0$, and expanding $\Phi$ (not necessarily real)  in powers of $s$, $\Phi(x,x',s)=\Phi_0(x,x') + \Phi_1(x,x')s + \ldots$, higher orders of $\Phi_k$ can recursively be computed with $\Phi_0$ alone incorporating the initial condition  \eqref{init_condition} in the form $\Phi_0(x',x')=0$ (note the manifest gauge covariance of this scheme to any order $k$).
For our purpose, $\Phi_0$ is enough. A simple calculation gives the gauge covariant phase
\beq
\Phi_0(x,x')=q\int_{x'}^x\d\xi\cdot A(\xi)\, ,
\enq  
where the integral is taken along the straight path connecting $x'$ with $x$. 

Focusing first on \eqref{integral_delta_equation}, we see that, for fixed $\gamma$ and $G$, it is in fact an equation for a function $f^\text{R}(s)\equiv\phi(\gamma_s,s)$. Indeed, plugging an ansatz for $f^\text{R}$ into the r.h.s. of \eqref{integral_delta_equation}, one can compute $\phi(x,s)$ $\forall s,x$, and in particular for $x=\gamma_s$, which we call $f^\text{L}(s)$. The linear map $f^\text{R}\mapsto f^\text{L}$ (which, using $G(x',x;s)=G^*(x,x';-s)$, can be shown to be formally self-adjoint) must therefore send $f^\text{R}$ to itself, for \eqref{integral_delta_equation} to have a solution. Now, the universal, viz. $A$-independent,  $i/ (2\pi\bar{h}s)^2$ divergence of $G(y,y,s)$ for $s\rightarrow 0$ and any $y$, implies $f^\text{R}\mapsto f^\text{R}+O(\epsilon)$, so the nontrivial content of \eqref{integral_delta_equation} is  in this $O(\epsilon)$ term, which we write as $\epsilon f^\text{r}$. In \cite{KY}, $\lim_{\epsilon\rightarrow 0}f^\text{r}=0$ was implied as the content of \eqref{integral_delta_equation}. While this may turn out to be true for some specific solutions (a freely moving particle, for example), 
equation \eqref{integral_delta_equation}  should take a more relaxed form
\beq\label{Im}
\text{Im } \left(\lim_{\epsilon \rightarrow 0}{f^\text{r}}^*\right)\,f^\text{R}=0\,,
\enq 
where, as usual, `Im' is the imaginary part of the entire product to its right.

Moving next  to the second ECD equation, \eqref{surfing_equation}, conveniently rewritten as 
\beq\label{surfing2}
\text{Re }\bar{h}\partial_x\phi(\gamma_s,s)\phi^*(\gamma_s,s)=0\,,
\enq
a similar isolation of the nontrivial content exists. For further use, however, we  first want to isolate the contribution  of the small $s$  divergence of $G$ to $\phi(x,s)$, for a general $x$ other than $\gamma_s$.  Substituting \eqref{short_s_propagator} into \eqref{integral_delta_equation}, and expanding the integrand around $s$ to first order in $s'-s$:
 $\gamma_{s'}\sim \gamma_s + \dot{\gamma}_s (s'-s)$,  $\Phi_0(x,\gamma_{s'})\sim\Phi_0(x,\gamma_s)$, $\phi(\gamma_{s'},s')\sim f^\text{R}(s)$, leads to a gauge covariant  definition of the \emph{singular part of $\phi$}  
\beq\label{divergent_phi}
\phi^\text{s}(x,s)= f^\text{R}(s)e^{i\big(\Phi_0(x,\gamma_s) +  \dot{\gamma}_s\cdot\xi\big)/\bar{h}}\, \text{sinc}\left(\frac{\xi^2}{2\bar{h}\epsilon}\right)\, 
\enq
with $\xi\equiv x-\gamma_s$.
Consequently, the \emph{residual} (or \emph{regular}) wave-function is defined via the gauge covariant equation
\beq\label{residual_phi}
\epsilon\phi^\text{r}(x,s)=\phi(x,s)-\phi^\text{s}(x,s)\,.
\enq
Note the implicit $\epsilon$-dependence of all the $\phi$'s in \eqref{residual_phi} which are omitted for economical reasons.
 
Using $\partial_x\Phi_0(x,\gamma_s)|_{x=\gamma_s}=qA(\gamma_s)$, we have
\beq\label{kj}
\phi^\text{s}(\gamma_s,s)=f^\text{R}(s)\,,\quad\bar{h}\partial_x\phi^\text{s}(\gamma_s,s)=i\big[\dot{\gamma}_s + A(\gamma_s)\big]f^\text{R}(s)
\,,
\enq
and \eqref{surfing2} is automatically satisfied up to an $O(\epsilon)$, gauge invariant term 
\beq\label{fgd}
 \epsilon\,\text{Re }\bar{h}\partial_x\big[\phi^\text{r}(\gamma_s,s) \phi^\text{s}(\gamma_s,s)^*\big]=\epsilon\,\text{Re }D\phi^\text{r}(\gamma_s,s) \phi^\text{s}(\gamma_s,s)^*\,,
\enq
where the above equality follows from \eqref{kj}, $\phi^\text{r}(\gamma_s,s)=f^\text{r}(s)$ and \eqref{Im}. The fine-tuned  definition of \eqref{surfing_equation} is therefore  
\beq\label{Re2}
\lim_{\epsilon\rightarrow 0}  \,\text{Re }D\phi^\text{r}(\gamma_s,s) \phi^\text{s}(\gamma_s,s)^*=0\,.
\enq
Using the above definitions, \eqref{Im} can also be written as 
\beq\label{Im2}
\lim_{\epsilon\rightarrow 0}\,\text{Im } \phi^\text{r}(\gamma_s,s)\phi^\text{s}(\gamma_s,s)^*=0\,.
\enq

More insight into this  fine tuned central ECD system is given in the sequel. For the time being, let us just note that it is invariant under the  original symmetry group of ECD. In particular, the system is invariant under 
\beq\label{stretch}
\phi^\text{s}\mapsto C\phi^\text{s}\,, \quad \phi^\text{r}\mapsto C\phi^\text{r}\,,\quad C\in{\mathbb C}\,,
\enq   
under a gauge transformation
\beq
A\mapsto A+\partial\Lambda\,,\quad G(x,x',s)\mapsto G\,e^{i\,\left[q\Lambda(x)-q\Lambda(x')\right]/\bar{h}}\,, \quad \phi^\text{s}\mapsto \phi^\text{s}e^{iq\Lambda/\bar{h}}\,, \quad \phi^\text{r}\mapsto \phi^\text{r}e^{iq\Lambda/\bar{h}}\,,
\enq
and under scaling of space-time
\begin {align}\label{scaling1}
&A(x)\mapsto\lambda^{-1} A(\lambda^{-1}x)\,,\quad\epsilon \mapsto \lambda^{2} \epsilon\,,\quad\gamma(s)\mapsto \lambda\gamma\left(\lambda^{-2}s\right)\,,\nonumber\\
&\phi^\text{s}\left(x,s\right)\mapsto \phi^\text{s}\left(\lambda^{-1} x,\lambda^{-2}s\right)\,,\quad\phi^\text{r}\left(x,s\right)\mapsto \lambda^{-2}\phi^\text{r}\left(\lambda^{-1} x,\lambda^{-2}s\right)\,,
\end {align}
directly following from the transformation of the propagator under scaling
\[
A(x)\mapsto\lambda^{-1} A(\lambda^{-1}x)\;\Rightarrow\; G(x,x';s)\mapsto\lambda^{-4} G\left(\lambda^{-1}x, \lambda^{-1} x'; \lambda^{-2} s\right)\,. 
\]
Regarding this  last symmetry, two points should be noted. First, for a finite $\epsilon$ it relates between solutions of  \emph{different} theories, indexed by different values of $\epsilon$. It is only because $\epsilon$ is ultimately eliminated from all results, via an $\epsilon\rightarrow 0$ limit, that scaling can be considered a symmetry of ECD. The second point concerns the scaling dimension of $\phi^\text{s}$ (and $\phi^\text{r}$). By the symmetry \eqref{stretch}, this dimension can be an arbitrary number $D$ ($D-2$ respectively). However, to comply with scale covariance $j$ must have dimension $-3$, hence $D=0$.

\subsection{ECD currents}\label{ECD_currents}
All ECD currents have the common form
\beq
j=\partial_\epsilon \epsilon^{-1} \int\d s\, B[\phi,\phi^*]\,,
\enq
where $B$ is some bilinear in $\phi$ and $\phi^*$. Using the decomposition \eqref{residual_phi} we write 
\beq
B[\phi,\phi^*]=\sum_{\text{a},\text{b}\in\{\text{r},\text{s}\}} O_\text{a} \phi^\text{a} \,O_\text{b}\phi^*{}^\text{b}\,,
\enq
for some local operators $O$'s (containing an $\epsilon$ multiplier in the case of r). There are therefore three types of contributions: $\{\text{a},\text{b}\}=\{\text{s},\text{s}\}, \{\text{r},\text{r}\}$, and $\{\text{r},\text{s}\}, \{\text{s},\text{r}\}$ taken as one. Let us examine each for the typical case of the electric current $j$---\eqref{ECD_current}.

The $\{\text{s},\text{s}\}$ term reads
\beq\label{ss}
j^\text{ss}(x)=\partial_\epsilon\epsilon^{-1}\frac{q}{\bar{h}}\int\d s\,\big(\dot{\gamma}_s - qA(x) + \partial_x\Phi(x,\gamma_s)\big)\left|f^\text{R}(s)\right|^2\,\text{sinc}^2\left(\frac{(x-\gamma_s)^2}{2\bar{h}\epsilon}\right)\, .
\enq 
By the same arguments as in section \ref{EC}, $j^\text{ss}(x)$ can be shown to reduce to the  line current
\beq\label{explicit_ss}
\alpha\int\d s\, \left|f^\text{R}(s)\right|^2 \delta^{(4)}\left(x-\gamma_s\right)\dot{\gamma}_s\,,
\enq
for some real constant $\alpha$, which is not necessarily conserved as  $\left|f^\text{R}(s)\right|^2$ may be $s$-dependent, and is discarded of in ECD. Likewise, the \{s,s\} contribution of all ECD currents is a distribution supported on $\bar{\gamma}$ albeit  generally containing more complex distributions, involving also derivatives of line distributions. 

Moving to the $\{\text{r},\text{r}\}$ term, this piece gives a nonsingular contribution which is well localized around $\bar{\gamma}$ in a region referred to as the core. The localization mechanism of the core is explained within the semiclassical approximation in appendix \ref{The classical limit}.

Finally,  the most singular part of the  $\{\text{r},\text{s}\}$ term gives  a regular, well localized piece, coming from the r part, multiplying an $x$-derivative of a $\delta\left(\xi^2\right)$ coming from the s part (in combination with the $\epsilon^{-1}$). This piece generates, at worst, an integrable $r^{-2}$ singularity on $\bar{\gamma}$ to which no charge leaks by virtue of \eqref{integral_delta_equation} (see appendix \ref{Conservation of ECD currents}). We say `at worst' because condition \eqref{Im2} could counter the $r^{-2}$ divergence coming from the delta-function, depending on the form  of $\phi^\text{r}(x,s)$ around $\gamma_s$. Indeed, in a semiclassical analysis the divergence is moderated to $r^{-1}$ only.

An $r^{-2}$ divergence in the $j^0$ component implies an $r^{-1}$ divergence of the electric field on $\bar{\gamma}$ which, in turn, leads to an integrable $r^{-2}$ divergence of $\Theta^{00}$---the electrostatic energy density (as oppose to the non-integrable $r^{-4}$ divergence in the case of a point charge.). Similar arguments show that $m^{00}$ has an integrable singularity at worst.

\section{A semiclassical analysis of ECD}\label{The classical limit}

In this section we analyze the consistency of the central ECD system using a small $\bar{h}$ approximation of the propagator known as the \emph{semiclassical propagator}, 
\begin {equation}\label{semi_classical_propagator}
G_\text{sc}(x,x';s)=\frac{i\, \text{sign}(s)}{(2 \pi \bar{h})^{2}}
{\cal F}(x,x';s) e^{i I(x,x';s)/\bar{h}} \, .
\end {equation}
Above,  
\beq\label{action}
\quad I = \int_{0}^{s}\text{d}\sigma\, \frac{1}{2}\dot{\beta}_\sigma^2 + qA(\beta_\sigma)\cdot\dot{\beta}_\sigma\,,
\enq
is the action of the\footnote{We shall assume that for a given $s$, there exists a unique  path connecting $x'$ with $x$. The existence of a plurality of classical paths is inconsequential to establishing the semiclassical limit. In Appendix B of \cite{KY} we analyze what `becomes' of such contributions for a finite ${\bar h}$, showing that they endow ECD particles with a `remote sensing mechanism'. } classical path $\beta$ satisfying $\beta_0=x'$ and $\beta_s=x$, and ${\cal F}$ --- the so-called Van-Vleck determinant --- is the gauge-invariant classical quantity, given by the determinant
\begin {equation} \label{Van Vleck}
{\cal F}(x,x';s)=
\left|\partial_{x_\mu}\partial_{x'_\nu}I(x,x';s)\right|^{1/2}\, .
\end {equation}

The semiclassical propagator becomes exact for small $s$, so  the singular-regular decomposition \eqref{residual_phi} of $\phi$ is consistent with the approximation, the latter affecting only the accuracy of $\phi^\text{r}$.\;\footnote{The approximation involved in the computation of the semiclassical propagator amounts to ignoring a `quantum potential' term in the dynamics of a classical particle originating from $x'$. This potential reads $\bar{h}^2\Box R/2R$, with $R$ the modulus of the exact propagator. Granted that the latter's form is \eqref{short_s_propagator} for small $s$, the modulus of $G$ is independent of $x$ and the quantum potential vanishes. }

Let us next show that to leading order in $\bar{h}$ and some fixed potential $A$, the fine-tuned central ECD system is solved by any classical $\gamma$, and by a corresponding   ansatz of the form 
\beq\label{ans1}
f^\text{R}(s')=Ce^{iI\left(\gamma_{s'},\gamma_0,s'\right)/\bar{h}}\,,
\enq
with $C\in{\mathbb C}$ an arbitrary constant.

Substituting in \eqref{integral_delta_equation}, $G\mapsto G_\text{sc}$, $x'\mapsto \gamma_{s'}$ and $x\mapsto \gamma_{s}$, we first note that  $\gamma$ is the classical path in $A$, connecting $\gamma_{s'}$ with $\gamma_s$.  Using 
\beq\label{sk}
I\left(\gamma_s,\gamma_{s'},s-s' \right)+ I\left(\gamma_{s'},\gamma_0,s' \right)=I\left(\gamma_s,\gamma_0,s \right)
\enq  
we get
\begin{align}\label{sd}
\phi(\gamma_s,s)&= \frac{\epsilon C}{2}e^{iI\left(\gamma_s,\gamma_0,s \right)/\bar{h}}\, \int_{-\infty}^{\infty}\text{d}s'\, 
{\cal F}\left(\gamma_s,\gamma_{s'};s-s'\right)\text{sign}(s-s')  {\mathcal
  U}(\epsilon;s-s')\nonumber \\
  &\Rightarrow \phi^\text{r}(\gamma_s,s)=\frac{ C}{2}e^{iI\left(\gamma_s,\gamma_0,s \right)/\bar{h}}\, \left[ R(s,\epsilon)-\frac{2}{\epsilon}\right]=\half f^\text{R}(s)\, \left[ R(s,\epsilon)-\frac{2}{\epsilon}\right]\, ,
\end{align} 
with
\beq\label{R}
R(s,\epsilon)=\int_{-\infty}^{\infty}\text{d}s'\, 
{\cal F}\left(\gamma_s,\gamma_{s'};s-s'\right)\text{sign}(s-s')  {\mathcal
  U}(\epsilon;s-s')
\enq 
some real functional of the EM field and its first derivative (its local neighborhood in an exact analysis) on $\bar{\gamma}$, such that $\lim_{\epsilon\rightarrow 0} \left[ R(s,\epsilon)-2/\epsilon\right]$ is finite, implying that \eqref{Im2} is satisfied.
  
Moving next to the second refined ECD equation, \eqref{Re2}, and pushing $\partial$ into the integral in \eqref{integral_delta_equation}, 
\begin{align}\label{derivative integral}
 &\bar{h}\partial \phi(\gamma_s,s) = \frac{\epsilon C}{2}e^{iI\left(\gamma_s,\gamma_0,s \right)/\bar{h}}\, \int_{-\infty}^{\infty} \text{d}s'\,
 \bigg[ i \partial_x I\left(x,\gamma_{s'};s-s'\right)\big|_{x=\gamma_s} \,
 {\cal F}(\gamma_s,\gamma_{s'};s-s')\\
                               &\quad+     \bar{h}\partial_x{\cal F}(x,\gamma_{s'};s-s')\big|_{x=\gamma_s}\bigg]
                                  \text{sign}(s-s')
{\cal U}(\epsilon;s-s')\,.\nonumber
\end{align}
The $\bar{h}\partial F$  term in \eqref{derivative integral} can be neglected for small $\bar{h}$.
Using a relativistic variant of the Hamilton-Jacobi theory (see appendix B  in \cite{KY}),
we can write  
\beq\label{4f}
\partial_x I\left(\gamma_s,\gamma_{s'},s-s'\right)=p(s)\equiv \dot{\gamma}_s + qA(\gamma_s)
\enq 
which is independent of $s'$. Together with \eqref{sd} we therefore get 
\begin {align}\label{vb}
&\bar{h}\partial \phi(\gamma_s,s)=ip(s)\phi(\gamma_s,s) \quad \Rightarrow\quad\bar{h}\partial \phi^\text{r}(\gamma_s,s)=ip(s)\phi^\text{r}(\gamma_s,s)\nonumber\\
&\Rightarrow\quad \lim_{\epsilon\rightarrow 0}\text{Re }D\phi^\text{r}(\gamma_s,s){f^\text{R}}^*(s)=-\dot{\gamma}_s \lim_{\epsilon\rightarrow 0}\text{Im } \phi^\text{r}(\gamma_s,s){f^\text{R}}^*(s)\,,
\end {align}
which vanishes by \eqref{Im2}, hence \eqref{Re2} is satisfied.

\subsection{ECD currents in the semiclassical approximation}
For $x$ other than $\gamma_s$, applying the semiclassical approximation to \eqref{integral_delta_equation} gives
\begin{align}\nonumber
\phi(x,s)&= \frac{\epsilon C}{2}\, \int_{-\infty}^{\infty}\text{d}s'\, 
{\cal F}\left(x,\gamma_{s'};s-s'\right)e^{i\left[I\left(x,\gamma_{s'},s-s' \right)+ I\left(\gamma_{s'},\gamma_0,s' \right)\right]/\bar{h}}\text{sign}(s-s')  {\mathcal
  U}(\epsilon;s-s')\,.
\end{align} 
The phase of the integrand is independent of $s'$ only for $x=\gamma_s$, as manifested in \eqref{sk}. Otherwise, the family of paths connecting $\gamma_{s'}$ with $x$, and that connecting $\gamma_{s'}$ with $\gamma_0$, traverse different parts of the potential and do not even lie on the same mass-shell. The phase is therefore  a rapidly oscillating function of $s'$ for small $\bar{h}$ and/or $x$ lying far from $\gamma_s$, rendering $\phi(x,s)$ arbitrarily localized around $\gamma_s$ in the limit $\bar{h}\rightarrow 0$. Combined with \eqref{4f} and a suitably chosen $C$, the ECD electric current \eqref{ECD_current} reduces to the CE electric current \eqref{classical_current} in that limit.  

In the  terminology of section \ref{ECD_currents}, using the accuracy of the semiclassical propagator for small $s$, it can readily be shown that the generic  integrable  singularity  of the $\{r,s\}$ contribution to  the ECD electric current \eqref{ECD_current} survives the semiclassical approximation. This singularity in $j^0$, moderated to $r^{-1}$ by virtue of \eqref{4f},  implies a discontinuous EM field at $\bar{\gamma}$ (a non differentiable $A$ there) which means that the fully coupled Maxwell-ECD system cannot be consistently solved in the semiclassical approximation as for  $x$ and $x'$  both lying on $\bar{\gamma}$, the semiclassical propagator  \eqref{semi_classical_propagator} is ill defined when self fields are taken into account (Note that this sensitivity to a discontinuity in the EM field is just an artifact of the semicalassical approximation and does not carry to an exact analysis.). 

In summary, the semiclassical analysis of (scalar) ECD has lead us back to the two well defined but mutually incompatible ingredients of CE: the Lorentz force equation and the line current associated with a point charge. We see once more that a solution to the classical self force problem requires an essentially ``quantum" ($\bar{h}\neq0$) treatment.

\section{The basic tenets}\label{Conservation of ECD currents}
To prove the conservation of the ECD electric current \eqref{ECD_current},  we first need the following lemma,
whose proof is obtained by direct computation.

\medskip

\noindent {\bf Lemma}. Let $f(x,s)$ and $g(x,s)$ be any (not
necessarily square integrable) two solutions of the homogeneous
Schr\"{o}dinger equation \eqref{Schrodinger}, then
\begin {equation}\label{fgstar}
\frac{\partial}{\partial s}(f g^*)=\partial_\mu \left[
\frac{i}{2}\big(D^\mu f g^* - \left(D^\mu g\right)^* f\big)
\right]\, .
\end {equation}

\medskip

\noindent This lemma is just a differential manifestation of
unitarity of the Schr\"{o}dinger evolution---hence the divergence.

Turning now to equation \eqref{integral_delta_equation},
\begin {equation}\label{integral_delta_equation2}
\phi(x,s)=  -2\pi^2\bar{h}^2 \epsilon i  \int_{-\infty}^{\infty}\text{d}s'\,
  \,G(x,\gamma_{s'};s-s') f^\text{R}(s'){\mathcal
  U}(\epsilon;s-s')\, ,
\end {equation}
and its complex conjugate,
\begin {equation}\label{integral_delta_equation2star}
\phi^*(x,s)=  2\pi^2\bar{h}^2 \epsilon i  \int_{-\infty}^{\infty}\text{d}s''\,
  \,G^*(x,\gamma_{s'};s-s'') {f^\text{R}}^*(s''){\mathcal
  U}(\epsilon;s-s'')\, ,
\end {equation}
we get by direct differentiation
\begin{align}\label{two terms}
&q\frac{\partial}{\partial s}\,\bigg[ -2\pi^2\bar{h}^2\epsilon i
\int_{-\infty}^{\infty}\text{d}s'f^\text{R}(s')\,\,\,2\pi^2\bar{h}^2  \epsilon i \int_{-\infty}^{\infty}\text{d}s''{f^\text{R}}^*(s'')\\
\nonumber
  &\qquad\qquad \qquad{\mathcal
  U}(\epsilon;s-s')G(x,\gamma_{s'};s-s')\,{\mathcal
  U}(\epsilon;s-s'')G^*(x,\gamma_{s''};s-s'')\,
  \bigg]
 \\ \nonumber
   &=
-2q\pi^2\bar{h}^2  \epsilon i
\int_{-\infty}^{\infty}\text{d}s'f^\text{R}(s')\,\,\,2\pi^2\bar{h}^2\epsilon i \int_{-\infty}^{\infty}\text{d}s''{f^\text{R}}^*(s'')\\
\nonumber &\qquad
\partial_s\big[G(x,\gamma_{s'};s-s')\,G^*(x,\gamma_{s''};s-s'')
  \big]{\mathcal
  U}(\epsilon;s-s'){\mathcal
  U}(\epsilon;s-s'')\\ \nonumber
  & + \quad\big[\partial_s {\mathcal
  U}(\epsilon;s-s'){\mathcal
  U}(\epsilon;s-s'') + {\mathcal
  U}(\epsilon;s-s')\partial_s {\mathcal
  U}(\epsilon;s-s'')\big]\\ \nonumber
 &\qquad \quad G(x,\gamma_{s'};s-s')\,G^*(x,\gamma_{s''};s-s'')\,
 .
\end{align}

\noindent Focusing on the first term on the r.h.s. of \eqref{two terms}, we note that, as $G$
is a homogeneous solution of Schr\"{o}dinger's equation, we can
apply our lemma to that term, which therefore reads
\begin{align}
&-2q\pi^2\bar{h}^2   \epsilon i\int_{-\infty}^{\infty}\text{d}s'f^\text{R}(s')\,\,\,2\pi^2\bar{h}^2 \epsilon i
\int_{-\infty}^{\infty}\text{d}s''{f^\text{R}}^*(s'')\\\nonumber
& \partial_\mu \left[ \frac{i}{2}\big(D^\mu G(x,\gamma_{s'};s-s')
G^*(x,\gamma_{s''};s-s'') - \big(D^\mu
G(x,\gamma_{s''};s-s'')\big)^* G(x,\gamma_{s'};s-s')\big) \right]\\
\nonumber &{\mathcal
  U}(\epsilon;s-s'){\mathcal
  U}(\epsilon;s-s'')\, .
\end{align} Integrating \eqref{two terms} with respect to $s$,
the left-hand side vanishes (we can safely assume it goes to zero
for all $x, s',s''$ as $|s|\rightarrow \infty$), and the
derivative $\partial_\mu$ can be pulled out of the triple
integral in the first term. The reader can verify that this  triple integral, after application of $\lim_{\epsilon\rightarrow 0} \partial_\epsilon \epsilon^{-1}$, is just
$\partial_\mu j^\mu$, with $j$ given by \eqref{ECD_current} and
$\phi$, $\phi^*$ are explicated using
\eqref{integral_delta_equation2},
\eqref{integral_delta_equation2star} respectively. The ECD electric current is therefore
conserved, provided the $s$ integral over the second term in \eqref{two
terms}, after application of $\lim_{\epsilon\rightarrow 0} \partial_\epsilon \epsilon^{-1}$ to it, vanishes in the distributional sense.

Let us then show that this is indeed the case. Integrating the second
term with respect to $s$, and using\\ $\partial_s {\mathcal
  U}(\epsilon;s-s') = \delta(s-s' - \epsilon)   + \delta(s -s'
  +\epsilon)$, that term reads
\begin{align}\label{second term}
&-2q\pi^2\bar{h}^2 \epsilon i
\int_{-\infty}^{\infty}\text{d}s'f^\text{R}(s')\,\,\,2\pi^2\bar{h}^2 \epsilon i \int_{-\infty}^{\infty}\text{d}s''{f^\text{R}}^*(s'')\, \\
\nonumber & {\mathcal
  U}(\epsilon;s'- \epsilon - s'')
  G(x,\gamma_{s'};-\epsilon)G^*(x,\gamma_{s''};s'-
  \epsilon-s'')\\ \nonumber
+&{\mathcal U}(\epsilon;s'+ \epsilon - s'')
  G(x,\gamma_{s'};+\epsilon)G^*(x,\gamma_{s''};s'+
  \epsilon-s'')\\ \nonumber
  +&{\mathcal U}(\epsilon;s''- \epsilon - s')
  G(x,\gamma_{s'};s''-\epsilon -s')G^*(x,\gamma_{s''};-\epsilon)\\ \nonumber
   +&{\mathcal U}(\epsilon;s''+ \epsilon - s')
  G(x,\gamma_{s'};s''+\epsilon -s')G^*(x,\gamma_{s''};+\epsilon)\, . \nonumber
\end{align}
Using \eqref{integral_delta_equation2} and
\eqref{integral_delta_equation2star}, this becomes
\begin{equation}\label{sum4}
 \text{Re}\,-4q\pi^2\bar{h}^2\epsilon i
  \int_{-\infty}^{\infty}\text{d}s'f^\text{R}(s')\Big[
\phi^*(x,s'-\epsilon)G(x,\gamma_{s'};-\epsilon) +\phi^*(x,s'+\epsilon)G(x,\gamma_{s'};\epsilon)\Big]\,.
\end{equation}
Writing $\phi=\phi^\text{s} + \epsilon\phi^\text{r}$ above, and using the short-$s$ propagator \eqref{short_s_propagator} plus the explicit form, \eqref{divergent_phi}, of $\phi^{s}$, one can show that  application of $\lim_{\epsilon\rightarrow 0} \partial_\epsilon \epsilon^{-1}$ to   \eqref{sum4} results in a distribution supported on $\bar{\gamma}$---a `line sink'---which is composed of two pieces: one coming from $\phi^\text{s}$ and one---from $\phi^\text{r}$. The s piece is just  the (not necessarily vanishing) divergence of the line current \eqref{explicit_ss} and is therefore of no concern to us. The second piece reads  
\begin{align}\label{dn}
 &\lim_{\epsilon\rightarrow 0}\,-8q\pi^2\bar{h}^2 
  \int_{-\infty}^{\infty}\text{d}s\, \text{Re }\,i\, f^\text{R}(s){\phi^\text{r}}^*(\gamma_s,s)\delta^{(4)}(x-\gamma_s)=\nonumber\\ &\lim_{\epsilon\rightarrow 0}\,8q\pi^2\bar{h}^2 
  \int_{-\infty}^{\infty}\text{d}s\, \text{Im }\, f^\text{R}(s){\phi^\text{r}}^*(\gamma_s,s)\delta^{(4)}(x-\gamma_s)\,
\end{align} 
and represents a `line sink in Minkowski's space' associated with the singularity of $j$ on $\bar{\gamma}$. By virtue of \eqref{Im2}, no leakage of charge occurs at those sinks, as one can  establish the time-independence of the charge by integrating $\partial\cdot j=0$ over a volume in Minkowski's space, and apply Stoke's theorem, to get a conserved quantity. A more explicit way of demonstrating the conservation of charge, avoiding the use of distributions, is shown next.

\subsection{Line sinks in Minkowski's space}\label{ls}
To gain a more  explicit geometrical  insight into the meaning of a `line sink in Minkowski's space', consider  a small space-like three-tube, $T$, surrounding $\bar{\gamma}$, the construction of which proceeds as follows. 
Let  $\beta(\tau)=\gamma\left(s(\tau)\right)$ be the world line $\bar{\gamma}$, parametrized by proper time $\tau=\int^s \sqrt{ (\d \gamma)^2 }$, and  let $x\mapsto {\tau_\text{r}}$ be the \emph{retarded light-cone map} defined by the relations
\beq\label{dh}
\eta^2\equiv \left(x-\beta_{\tau_\text{r}}\right)^2=0\,,\quad \text{and}\quad \eta^0>0\,.
\enq
Let  the `retarded radius' of $x$ be
\beq
r=\eta\cdot\dot{\beta}_{\tau_\text{r}}\,.
\enq
Taking the derivative of \eqref{dh}, treating ${\tau_\text{r}}$ as an implicit function of $x$, and solving for $\partial{\tau_\text{r}}$, we get
\beq\label{yd}
\partial {\tau_\text{r}}=\frac{\eta}{r}\quad\Rightarrow\quad \partial r= \dot{\beta}_{\tau_\text{r}} + \left(\ddot{\beta}_{\tau_\text{r}}\cdot\eta -1\right)\frac{\eta}{r}\,.
\enq 
The (retarded) three-tube of radius $\rho$ is defined as the time-like three surface 
\[
T_\rho=\left\{x\in {\mathbb M}:\, r(x)=\rho\right\}\,.
\]
It can be shown in a standard way that the  directed surface element normal to $x\in T_\rho$ is 
\beq
\d^\mu T_\rho=\partial^\mu r\big|_{r=\rho}\, \rho^2\, \d \tau\,\d \Omega\,,
\enq
where $\d\Omega$ is the surface element on the two-sphere.

Let $\Sigma_1$ and $\Sigma_2$ be two space-like surfaces, intersecting $T_\rho$ and $T_R$. Applying Stoke's theorem to the interior of the three surface composed of $T_\rho$, $T_R$, $\Sigma_1$ and $\Sigma_2$, and using $\partial \cdot j=0$ there,  we get   
\beq\label{gb}
\int_{\Sigma_2}\d\Sigma_2\cdot j + \int_{\Sigma_1}\d\Sigma_1\cdot j=-\int_{T_\rho}\d T_\rho\cdot j-\int_{T_R}\d T_R\cdot j\,.
\enq
Realistically assuming that the second term on the r.h.s. of \eqref{gb} vanishes for $R\rightarrow \infty$, we get that the `leakage' of the charge, $\int_{\Sigma_2}\d\Sigma_2\cdot j - \int_{\Sigma_1}\d\Sigma_1\cdot j$, equals to $-\lim_{\rho\rightarrow 0}\int_{T_\rho}\d T_\rho\cdot j$.  

As $\d T_\rho=O(\rho^2)$, the leakage  only involves the piece of $j$ diverging as $r^{-2}$. This piece, reads  
\begin {align}
&2q\bar{h}^2\int\d s\, \text{Im }{\phi^\text{r}}^*(x,s)f^\text{R}(s)\partial \frac{1}{2\bar{h}\epsilon}\text{sinc}\left(\frac{\xi^2}{2\bar{h}\epsilon}\right)\;
\underset{\epsilon\rightarrow 0}{\longrightarrow}\;2q\bar{h}^2\pi\int\d s\, \text{Im }{\phi^\text{r}}^*(x,s)f^\text{R}(s)\partial\delta\left(\xi^2\right)\nonumber\\
&\sim\;2q\bar{h}^2\pi\, \partial\int\d s\, \text{Im }{\phi^\text{r}}^*(\gamma_s,s)f^\text{R}(s)\delta\left(\xi^2\right)\;=\;q\bar{h}^2\pi\sum_{s=s_\text{r}, s_\text{a}} \text{Im }{\phi^\text{r}}^*(\gamma_s,s)f^\text{R}(s)\, \partial\frac{1}{\left|\xi\cdot\dot{\gamma}_s\right|}\nonumber\,,
\end {align}
where $s_\text{r}=s\left({\tau_\text{r}})\right)$, and $\gamma_{s_\text{a}}$ is the corresponding \emph{advanced} point on $\bar{\gamma}$, defined by \[\xi^2\equiv \left(x-\gamma_{s_a}\right)^2 =0\,,\quad \xi^0<0\,.\] Focusing first on the contribution of $s_\text{r}$, and using a technique similar to that leading to \eqref{yd}, we get
\beq
\partial\frac{1}{\xi\cdot\dot{\gamma}_{s_\text{r}}}=-\frac{\dot{\gamma}_{s_\text{r}}}{\left(\xi\cdot\dot{\gamma}_{s_\text{r}}\right)^2}
+\frac{\left(\dot{\gamma}^2_{s_\text{r}} + \ddot{\gamma}_{s_\text{r}}\cdot\xi\right)\xi}{\left(\xi\cdot\dot{\gamma}_{s_\text{r}}\right)^3}\underset{\xi\rightarrow 0}{\sim}-\frac{\dot{\beta}_{\tau_\text{r}}}{mr^2} + \frac{\eta}{mr^3}\,,
\enq
where $m=\d\tau/\d s$ needs not be constant. In the limit $\rho\rightarrow 0$, using $\partial\frac{1}{\xi\cdot\dot{\gamma}_{s_\text{r}}}\cdot \partial r|_{r=\rho}\sim 1/(m\rho^2)$,  the contribution of $s_\text{r}$ to the flux across $T_\rho$  is most easily computed 
\begin {align}\label{gs}
&\int_{T_\rho}\d T_\rho\cdot j=q\bar{h}^2 \pi \int\d \Omega\int\d \tau_\text{r} m^{-1}\,\text{Im }{\phi^\text{r}}^*(\beta_{\tau_\text{r}},\tau_\text{r})f^\text{R}(\tau_\text{r})\nonumber\\
&=4q\bar{h}^2 \pi^2 \int\d s_\text{r}\,\text{Im }{\phi^\text{r}}^*(\beta_{s_\text{r}},s_\text{r})f^\text{R}(s_\text{r})\,.
\end {align}
The contribution of  $s_\text{a}$ to the flux  of $j$ is more easily computed   across a \emph{different}, (advanced) $T_\rho$, and gives  the same result in the limit $\rho\rightarrow 0$. The fact that $\rho$ can be taken arbitrarily small, in conjunction with the conservation of $j(x)$ for $x\notin \bar{\gamma}$, implies that the flux of $j$ across \emph{any} three-tube, $T=\partial C$, with $C$ a three-cylinder containing $\bar{\gamma}$, equals twice the value in \eqref{gs}, when $C$ is shrunk to $\bar{\gamma}$. Changing the dummy variable $s_\text{r}\mapsto s$ in \eqref{gs}, the formal  content of \eqref{dn} receives a clear  meaning using Stoke's theorem 
\[
\int_C\d^4 x\, \partial\cdot j= 8q\bar{h}^2 \pi^2 \int\d s\,\text{Im }{\phi^\text{r}}^*(\beta_{s},s)f^\text{R}(s)\int_C\d^4 x\,\delta^{(4)}(x-\gamma_s)=\int_T\d T\cdot j\,,
\]
which vanishes by virtue of \eqref{Im2}.

\subsection{Energy-momentum conservation}\label{Energy-momentum conservation}
The conservation of the ECD energy momentum tensor  can be established by the same technique used in the previous section. To explore yet another technique, as well as to illustrate the role played by symmetries of ECD in the context of conservation laws, we cautiously apply Noether's theorem to the  following functional 
\beq\label{A}
L\left[\left\{\lsuper{k}\varphi\right\}_1^n, {\cal A}\right]=\sum_{k=1}^n  L_{\text m}\left[\lsuper{k}\varphi,{\cal A}\right] - \int_{{\textsl M}}\d^4 x \frac{1}{4} {\cal F}^{\mu\nu}{\cal F}_{\mu\nu}
\enq
\beq\label{matter L}
\text{with}\qquad L_{\text m}[\varphi,{\cal A}]=-\int_{-\infty}^\infty\d s\int_{{\textsl M}}\d^4 x
\frac{i \bar{h}}{2}\left(  \varphi^*\partial_s \varphi\,
 - \partial_s \varphi^* \varphi \right)   -
  \frac{1}{2}\left(D^\lambda \varphi\right)^* D_\lambda \varphi\, .
\enq
Above, ${\cal A}$ is a vector-potential which appears in the covariant derivative, $D$,  ${\cal F}$ the associated Faraday tensor, and $\varphi$ a complex scalar function of space-time and of $s$.

Next, let us look at a solution to the coupled ECD-Maxwell system of $n$ interacting particles. As an $\epsilon$ limit is involved in that solution, we assume that in addition to a vector potential, $A$, we also also have a family of $n$ pairs, $ \left\{\lsuper{k}\phi_\epsilon,\lsuper{k}\gamma\right\}$, such that $\lim_{\epsilon\rightarrow 0} \epsilon^{-1}\,\lsuper{k}\phi_\epsilon$ exists in the distributional sense (the $\epsilon^{-1}$ is needed to render the limit non trivial).

Plugging $A$ and the form  \eqref{integral_delta_equation2} of $\phi_\epsilon$  into \eqref{A}, we compute the first variation of \eqref{A} around them: \footnote{We restrict the space of variations to those which go sufficiently fast to zero at infinity to justify integrations by part.}

\beq\label{delta_A}
\delta L=  \int_{-\infty}^\infty\d s\int_{{\textsl M}}\d^4 x \, -\left( \sum_k\,q\text{Im }\phi_\epsilon^*D^\mu\phi_\epsilon -  \partial_\nu F^{\nu \mu}\right)\delta A^\mu-\sum_k\,2\text{Re } \big(i\partial_s - {\cal H}\big)\phi_\epsilon\delta\phi_\epsilon^*
\enq
where ${\cal H}$ is defined in \eqref{Schrodinger}---without a scalar potential. Note that (obvious) particle labels, $k$, have been dropped  for economical reasons.

Choosing $\delta\phi_\epsilon = \partial \phi_\epsilon\cdot a$, and $\delta A^\mu=\partial_\nu A^\mu a^\nu$, corresponding to an infinitesimal shift, $x\mapsto x+a(x)$, of the coordinates, we get after some integrations by part 
\begin {align}\label{do}
&\delta L= \int_{{\textsl M}}\d^4 x\, \partial_\nu p^{\nu \mu}\, a_\mu\;\underset{\text{by eq. \eqref{delta_A}}}{=}- \left(\sum_k\, q\text{Im }\phi_\epsilon^*D^\mu\phi_\epsilon -  \partial_\nu F^{\nu \mu}\right)\partial_\nu A^\mu a^\nu-\sum_k\\
&\int_{-\infty}^\infty\d s\int_{{\textsl M}}\d^4 x\,\text{Re } 4\pi^2\bar{h}^2\epsilon \big[G\left(x,\gamma_{s-\epsilon};+\epsilon\right) f^\text{R}(s-\epsilon) + G\left(x,\gamma_{s+\epsilon};-\epsilon\right) f^\text{R}(s+\epsilon)\big]\partial^\mu \phi^*(x,s)\, a_\mu\nonumber \,,
\end {align}
with $p$ the Noether (canonical) e-m tensor associated with the action \eqref{A}, computed for $\phi_\epsilon$ and $A$, and the second line follows from
\beq
\big(i\partial_s - {\cal H}\big)\phi= 2\pi^2\bar{h}^2 \epsilon \bigg[G\left(x,\gamma_{s-\epsilon};+\epsilon\right) f^\text{R}(s-\epsilon) + G\left(x,\gamma_{s+\epsilon};-\epsilon\right) f^\text{R}(s+\epsilon)\bigg]\, ,
\enq   
which, in turn, directly follows from \eqref{integral_delta_equation2}.

Applying now $\lim_{\epsilon\rightarrow 0} \partial_\epsilon\epsilon^{-1}$ to \eqref{do}, the first term on the r.h.s. vanishes by virtue of Maxwell's equations and the definition, \eqref{ECD_current}, of the electric current, while the terms that follow can be analyzed using the same technique used in the computation of \eqref{sum4}. This gives
\begin{align}\label{fb}
 &8\pi^2\bar{h}^2 
 \int_{-\infty}^{\infty}\text{d}s\, \int_{{\textsl M}}\d^4 x\, \text{Re } f^\text{R}(s)\delta^{(4)}(x-\gamma_s)\partial{\phi^\text{r}}^*(\gamma_s,s)\cdot a(x,s)=\nonumber\\
 &8\pi^2\bar{h}^2 \int_{-\infty}^{\infty}\text{d}s\,\text{Re } f^\text{R}(s)\partial{\phi^\text{r}}^*(\gamma_s,s)\cdot a(\gamma_s,s)\,
\end{align}
which vanishes by virtue of \eqref{Re2} \emph{for any} $a$. The arbitrariness of $a$ implies the vanishing of  $\partial_\nu\left( \lim_{\epsilon\rightarrow 0} \partial_\epsilon\epsilon^{-1} p^{\nu \mu}\right)$ in the distributional sense.
Just like  the  electric current $j$, the  e-m tensor $p$  can easily be shown to be a smooth function of $x$, implying  point-wise conservation.  Equation \eqref{Re2} in the central ECD system, by which \eqref{fb} vanishes, appears therefore as the condition  that no  energy or momentum leak into a world-sink on $\bar{\gamma}$.

The Noether e-m tensor $p$ is not symmetric nor gauge invariant. This is an artefact of a non generally covariant treatment and the standard way of dealing with it (other than by setting a $g_{\mu\nu}=\eta_{\mu\nu}$ in the generally covariant version) is to add to $p$ a  conserved chargeless piece $\partial_\lambda\left(F^{\nu \lambda} A_\mu \right)$ which together, again using Maxwell's equations, turn $p$ into  $\Theta+\sum_k \lsuper{k}m$, with $\Theta$ the canonical tensor \eqref{Theta} and $\lsuper{k}m$ the `mechanical' e-m tensor \eqref{ECD_m}.

\subsection{Charges leaking  into world sinks}
Both methods used above, can be applied to prove the conservation of the   mass-squared current --- the counterpart of \eqref{msc}
\beq\label{b}
b(x)=\lim_{\epsilon\rightarrow 0} \partial_\epsilon\epsilon^{-1}\int\d s\,B(x,s)\equiv \lim_{\epsilon\rightarrow 0} \partial_\epsilon\epsilon^{-1}\int\d s\, \text{Re } \bar{h} \partial_s\phi^*D\phi\,,\quad
\text{for } x\notin \bar{\gamma}\,.
\enq
In the first method, used to establish the conservation of $j$, the counterpart of \eqref{fgstar} is  $\partial_s\left(g^*{\cal H}f\right) = \partial\cdot\left(\text{Re }\bar{h}\partial_s g^* Df\right)$, corresponding to the invariance of the Hamiltonian (in the Heisenberg picture) under the Schr{\"o}dinger evolution. In the variational approach, the conservation follows from the (formal) invariance of \eqref{A} $\phi(x,s)\mapsto \phi(x,s+s_0)$. However,  the leakage to the sink on $\bar{\gamma}$, between $\gamma_{s_1}$ and $\gamma_{s_2}$, is given by
\beq\label{jd}
8\pi^2\bar{h}^3 
  \int_{s_1}^{s_2}\text{d}s\, \text{Re }\,\partial_s{\phi^\text{r}}^*(\gamma_s,s) f^\text{R}(s)\,,
\enq   
is not guaranteed to vanish. Note that this leakage (whether positive or negative) is a `highly quantum' phenomenon --- proportional to $\bar{h}^2$ (the term $\partial_s{\phi^\text{r}}$ generally diverges as $\bar{h}^{-1}$).

Similarly, associated with the formal invariance of \eqref{A} under
\[
A(x)\mapsto\lambda^{-1}A\left(\lambda^{-1}x\right)\,,\quad\phi(x,s)\mapsto\lambda^{-2}\phi\left(\lambda^{-1}x,\lambda^{-2}s\right)\,,
\]
is a locally conserved dilatation current, the counterpart of the classical current \eqref{dc},
\begin {equation}\label{dilatation current}
 \xi^\mu= p^{\mu \nu} x_\nu -\lim_{\epsilon\rightarrow 0} \partial_\epsilon\epsilon^{-1}\sum_k 2\int_{-\infty}^{\infty}
 \d s\,  s\; \lsuper{k}B\, ,\quad\text{{\footnotesize with $B$ defined in \eqref{b}}}\,.
\end {equation}
At a large distance from an isolated particle, the term $p^{\mu \nu} x_\nu$ above  becomes $\Theta^{\mu \nu}x_\nu$ and can be shown to vanish. A possible change to a particle's scale-charge can only be due to leakage of the second term, involving the mass-squared of the particles, to its sink on $\lsuper{k}\bar{\gamma}$.  A leakage of mass, therefore, also modifies the scale-charge of a solution.

\section{The Lorentz force from the basic tenets}\label{single}

The derivation of the Lorentz force equation \eqref{Lorentz_force} given below, applies to a general, sufficiently isolated pair $j$, $m$, satisfying  the basic tenets and  co-localized---in any sensibly meaning of the word---around a world-line  $\bar{\gamma}=\cup_s \gamma_s$. Explicit assumptions are made which are natural in certain regimes yet utterly not in others, e.g. atomic scales, where the validity of the Lorentz force is in sharp conflict with experiment.    

\begin{figure}
	\centering
		\includegraphics[width=0.7\textwidth]{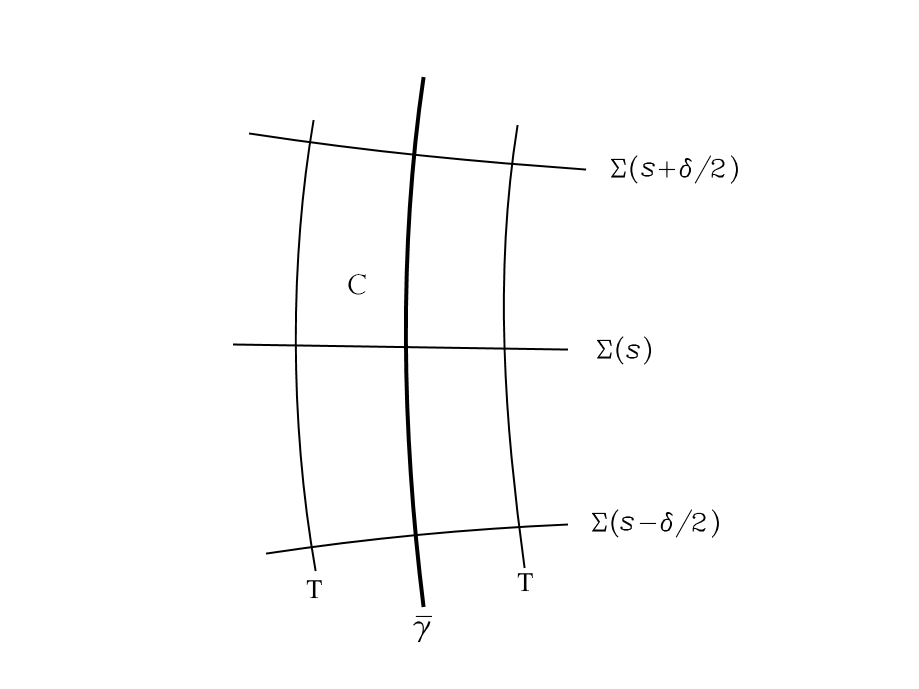}
		\caption{A $1+1$ spacetime  counterpart of our construction.}
	\label{fig:diagram}
\end{figure}

Let $\Sigma(s)$ be a foliation of ${\textsl M}$, viz., a one-parameter family of non intersecting space-like surfaces, each orthogonally intersecting the world line $\bar{\gamma}$  at $\gamma_s$,  $C$ a four-cylinder containing $\bar{\gamma}$, and $p^\mu(\tau)$  the corresponding four-momenta
\beq\label{jkc}
p^\mu(s)=\int_{\Sigma(s)\cap C}\d\Sigma_\nu\, m^{\nu\mu}\,,
\enq
where $\d\Sigma$ is the Lorentz covariant directed surface element, orthogonal to $\Sigma(s)$. Let also $C(s,\delta)\in C$ be the volume enclosed between $\Sigma(s-\delta/2)$ and $\Sigma(s+\delta/2)$, and $T(s,\delta)$ its time-like boundary (see figure \ref{fig:diagram}). Integrating \eqref{ni} over $C(s,\delta)$,   applying Stoke's theorem to the l.h.s., and dividing by $\delta$ we get
\beq\label{fka}
\frac{p^\mu(s+\delta/2) - p^\mu(s-\delta/2)}{\delta} + \delta^{-1}\int_T \d T_\nu\, m^{\nu \mu} = \delta^{-1}\int_{C(s,\delta)} \d^4 x F^{\mu \nu} j_\nu\,,
\enq
\noindent with $\d T$ the outward pointing directed surface element, orthogonal to  $T$. Assuming  that  $m$ is sufficiently localized about $\bar{\gamma}$,  the second, surface term, on the l.h.s. of \eqref{fka} can be ignored.

Both sides of \eqref{fka} depend on the details of the foliation $\{\Sigma_s\}$, and may rapidly fluctuate if the particle experiences internal vibrations. Both, nevertheless, are well defined---unlike in the point-charge case.

To translate \eqref{fka} into an equation for $\bar{\gamma}$---roughly speaking the center of the current---we first `low-pass' \eqref{fka}, viz.,  convolve it with a normalized kernel, $w(s)$, to remove possible fluctuations  which are due to internal vibrations in the particle. It is easy to see then that for a sufficiently wide $w$, the r.h.s. of \eqref{fka} becomes independent of the details of the foliation hence also the l.h.s. of \eqref{fka} (remember that we ignore the second term on the l.h.s.).  Next, we make the assumptions that the low-passed $p$ is locally ($s$-wise) proportional to the low-passed $\dot{\gamma}$, with an $s$-independent proportionality constant $M_m$.  This latter assumption is nothing but the condition that the same particle is being investigated at different $s$'s, namely, that the average mechanical momentum of the particle can be deduced from its average velocity.  We further assume that the low-passed momentum is slowly changing on time scales on the order of $\delta$. Under these assumptions,  using the same notation for the low-passed $\gamma$, \eqref{fka} becomes
\beq\label{kwjf}
M_m\ddot{\gamma}^\mu=\delta^{-1}\int \d^4 x\, \bar{w}(s,x) F^{\mu \nu}(x) j_\nu(x) \equiv   \left\langle F^{\mu \nu} j_\nu\  \right\rangle_{\gamma_s} 
\enq
with $\bar{w}(s,x)$ defined by $x\in\Sigma_{s'}\Rightarrow \bar{w}(s,x)=w(s-s')$, and $M_m=\sqrt{p^2}$ is the `mechanical mass'. This is a natural assumption for slowly varying, weak external fields. 

For a sufficiently isolated particle, expression \eqref{convolution with K} for $A$ provides a convenient decomposition of $F$ in \eqref{kwjf} into a self field, $F_\text{sel}$ generated by the isolated particle, and an external field $F_\text{ext}$ generated by the rest of the particles. For a slowly varying $F_\text{ext}$ on the scale set by $w$ the r.h.s. of \eqref{kwjf} can be written
\beq\label{jjd}
QF_\text{ext}{}^{\mu \nu}(\gamma_s)\dot{\gamma}_\nu + \left\langle F_\text{sel}{}^{\mu \nu} j_\nu\  \right\rangle_{\gamma_s}\,,
\enq
with $Q=\int_{\Sigma_s}\d\Sigma\cdot j$ the $s$-independent electric charge. 

The self-force term in \eqref{jjd} is dealt with by noting that $j$ generates $F_\text{sel}$, hence we can locally apply Poynting theorem \eqref{Poynting} to them, calling $\Theta_\text{sel}$ the associated canonical tensor. Further noting that \eqref{Poynting} is formally equivalent to \eqref{ni} used above with $m\mapsto -\Theta_\text{sel}$,  we can hope that, insofar as as the particle does not (significantly) radiate, the following would also be true 
\beq\label{hdia}
 \left\langle F_\text{sel}{}^{\mu \nu} j_\nu\  \right\rangle_{\gamma_s}=-M_\text{EM}\ddot{\gamma}\,,
\enq
with $M_\text{EM}=\sqrt{p_\text{EM}^2}$, where 
\beq\label{fhk}
p_\text{EM}^\mu=\int_{\Sigma(s)\cap C}\d\Sigma_\nu\, \Theta_\text{sel}^{\nu\mu}\,,
\enq
the EM contribution to a particle's e-m.  Combining \eqref{kwjf}, \eqref{jjd} and \eqref{hdia}, we get the Lorentz force equation for a particle with an effective mass equal to $M_m + M_\text{EM}$ when  $s$ is chosen as the proper time $\tau$ ($\d\tau= \sqrt{\d\bar{\gamma}^2}$). Note that
\beq\label{xdo}
 M_\text{EM}=Q^2\big(C/(2\epsilon)-1/(2\rho)\big)
\enq 
with $\epsilon$ the `extent' of $j^0$ the rest-frame of $\gamma$, defined up to some constant $C$ on the order of $1$, and $\rho$ the radius of the tube $T$. This requires $m$ to similarly diverge as $\sim -CQ^2/(2\epsilon)$ for the point  charge limit to have a well defined `physical' (observed) mass. However, when `absorbing' this way the Coulomb infinity into the physical mass, we are still left with a $\rho$-dependence in \eqref{xdo},  which is an absurd if the point-charge limit is to have a well defined meaning, independent of the arbitrary $T$ chosen to analyze e-m conservation. We therefore must calculate 
\beq\label{dfs}
\left\langle F_\text{sel}{}^{\mu \nu} j_\nu\  \right\rangle_{\gamma_s}=-\delta^{-1} \int_{\partial C(s,\delta)} \d C_\nu\, \Theta_\text{sel}^{\nu \mu} 
\enq
more prudently, with $\d C$ standing for both $\d T$ and $\d \Sigma$, beyond the sloppy approximation \eqref{fhk}. To this end,  one must commit to a particular mixture of advanced and retarded    solutions (which is incompatible with ECD's mathematical structure; recall section  \ref{FDM}). Adhering to the convention of using only the  retarded solution---the same is true for Dirac's choice of half-retarded-minus-half-advanced field---the integral \eqref{dfs} can be evaluated in the point-charge viz., $\epsilon\rightarrow 0$ and $\delta\rightarrow 0$ limit, further assuming that only  the dipole term associated with its bulk motion is responsible for the leading  contribution away from $j$. The result reads: $-CQ^2/(2\epsilon)\ddot{\gamma} + Q^2\frac{2}{3} (\ddot{\gamma}^2\dot{\gamma}+\dddot{\gamma})+O(\rho)$, where the $\rho\rightarrow 0$ limit is taken  \emph{after} the $\epsilon\rightarrow 0$ one.  We see that the $\rho$-dependence in \eqref{fhk} is canceled by an opposite  term coming from the integral over $T$.

A more symmetric treatment of `matter' and the EM field  is provided by the conservation of the total e-m, $p$ in \eqref{pp}.  Applying Stoke's theorem to $\partial p=0$, and using the same construction as in figure 1, we get
\beq\label{ppp}
p^\mu(s+\delta/2) - p^\mu(s-\delta/2) =-\int_T \d T_\nu\, p^{\nu \mu}\,,
\enq    
with 
\beq
p^\mu(s)=\int_{\Sigma(s)\cap C}\d\Sigma_\nu\, p^{\nu\mu}\,,
\enq
the total four-momentum content of $\Sigma(s)\cap C$, including the EM part coming from $\Theta$---the full canonical tensor this time. If we assume, as previously, that the flux of $p$ across $T$ is purely of EM origin, we arrive at the conclusion that, for a sufficiently isolated particle (or a bound aggregate of particles), the change in momentum can be read from the flux of the Poynting vector across a time-like surface surrounding it. Note that no approximation whatsoever is involved this time.

\subsection{ALD violates energy-momentum conservation}

Our `derivation' of the ALD equation requires a (covariant) regulator, $\epsilon$---the `size' of the charge---keeping the Coulomb divergence at bay, which is taken to zero only as a final stage. However, cancellation of this divergence in the $\epsilon\rightarrow 0$ limit by an opposite divergence in the mechanical e-m tensor, $m$, is just a necessary condition for a non trivial final result. As in section \ref{EC}, all the `action' gets squeezed to a line, and it is far from obvious that the basic tenets can `survive' this point-charge limit and, if they do, that all assumptions made in the current derivation of ALD, are compatible with this limit. In fact, they are almost certainly not, as the following example shows. Consider  two or more charges arriving from infinity, interacting, and then scattering back to infinity. If e-m is indeed conserved, then the sum of incoming momenta should equals that of outgoing, the latter comprising both mechanical and (scattered) EM waves.  
Integrating the ALD equation satisfied by any such charge,   the difference between its incoming and outgoing mechanical momentum, $m\dot{\gamma}\left|_{-\infty}^{\infty}\right.$, is indeed  due to radiation losses to infinity---the integral of $Q^2\frac{2}{3} \ddot{\gamma}^2\dot{\gamma}$---but there is also  the work of the Lorentz force due to all \emph{other} scattered charges (the integral of the $\dddot{\gamma}_s$ term, being a derivative, vanishes).  Now, unlike in Newtonian mechanics, two charges interacting through their retarded fields only, do not enjoy the property that, the force of particle 1 on particle 2 equals minus that of particle 2 on 1; The action and the reaction are due to two distinct forces, at distinct times, and their integrals need not cancel when summing over all charges. For global e-m to be conserved in such  scattering experiments, interference effects at infinity between the retarded fields generated by individual charges should compensate for the previous imbalance, but this is clearly impossible for an arbitrary scattering experiment, involving an arbitrary number of charges and masses (Yet another option salvaging e-m conservation is if the outgoing charges keep accelerating indefinitely---so-called runaway solutions---in which case the  $\dddot{\gamma}_s$ term can't be discarded, but this option is clearly non physical). The moral is that advanced fields must also play a role if global e-m conservation is to be satisfied, and this conclusion obviously applies also to the reduced order ALD equation or any other local correction to the Lorentz force equation.


\section{Spin-$\half$ ECD}\label{Spin-half}
In a spin-{\small $\half$}  version of ECD, the following modifications are made. The wave-function $\phi$ is a bispinor (${\mathbb C}^4$-valued), transforming in a Lorentz transformation according to 
\beq\label{hl}
\rho\left(e^{\omega}\right)\phi\equiv e^{-i/4\,\sigma_{\mu\nu}\omega^{\mu\nu}}\phi\, , \quad \text{for } e^{\omega}\in SO(3,1)\, , 
\enq
where $\sigma_{\mu \nu} =
\frac{i}{2}\left[\gamma_\mu, \gamma_\nu \right]$, with
$\gamma_\mu$  Dirac matrices (not to be confused with $\gamma$ the
trajectory).

The propagator is now a complex, $4\times4$ matrix, transforming under the adjoint representation, satisfying
\begin {equation}\label{spin_Schrodinger}
 i\bar{h}\partial_s G(x,x',s)=-\half{\slashed D}^2 G(x,x',s) \, ,\qquad {\slashed a}\equiv\gamma^\mu a_\mu\,,
\end {equation}
with the initial condition \eqref{init_condition} at
$s\rightarrow 0$  reading  $\delta^{(4)}(x-x')\delta_{\alpha\beta}$, where $\delta_{\alpha\beta}$ is the identity operator in spinor-space.

The transition to  spin-$\half$ ECD is rendered easy by the observation  that all expressions in scalar ECD are sums of bilinears of the form $a^*b$, which can be seen as a Lorentz invariant scalar product in ${\mathbb C}^1$. 
Defining an  inner product in spinor space (instead of ${\mathbb C}^1$)
\beq\label{inner_product}
(a,b)\equiv \,a^\dagger\gamma^0 b\, ,
\enq  
with $\gamma^0$  the Dirac matrix $\text{diag}(1,1,-1,-1)$ (again, not to be confused with $\gamma$ the trajectory) and substituting $a^*b\mapsto (a,b)$ in all bilinears, all the results of scalar ECD are retained. The Lorentz invariance of \eqref{inner_product} follows from  the Hermiticity of $\sigma^{\mu\nu}$ with respect to that inner product, viz. $\left(\sigma^{\mu\nu}\right)^\dagger=\gamma^0\sigma^{\mu\nu}\gamma^0$, and from $\left(\gamma^0\right)^2=1$.

Let us illustrate this procedure for important cases.
By a direct calculation of the short-$s$ propagator of \eqref{spin_Schrodinger}, as in section \ref{refined_ECD},  the spin can be show to affect the $O(s)$ terms in the expansion of $\Phi$, leading to an equally simple  $\phi^\text{s}$, the counterpart of \eqref{divergent_phi}, from which the regular part of all ECD currents can be obtained. The action, \eqref{matter L}, from which all conservation laws can be derived, is modified to 
\beq\label{A_s}
L_\text{m}[\varphi,{\cal A}]=\int_{-\infty}^\infty\d s\int_{{\mathbb M}}\d^4 x
\frac{i \bar{h}}{2}\big[  \left(\varphi,\partial_s \varphi\right)\,
 - \left(\partial_s \varphi, \varphi\right) \big]   -
  \frac{1}{2}\left({\slashed D} \varphi, {\slashed D} \varphi\right)\, ,
\enq 
while the counterpart of the electric current, \eqref{ECD_current}, derived from $\phi$, is now a sum of an `orbital current' and a `spin current'
\beq\label{jk}
j^\mu(x)\equiv j^\text{orb}{}^\mu + j^\text{spn}{}^\mu= \lim_{\epsilon\rightarrow 0} \partial_\epsilon\epsilon^{-1}\int\d s\; q\text{Im}\, (\phi,D^\mu\phi)\; - \bar{h}\partial_\nu(\phi,\sigma^{\nu \mu}\phi)\;{}\,, \quad \text{for } x\notin\bar{\gamma}\, .
\enq 
Each of the terms composing $j$ is individually conserved and gauge invariant. The conservation of the sum is follows from the $U(1)$ invariance of \eqref{A_s}, while conservation of the spin current follows directly from the antisymmetry of $\sigma$. This  current has an interesting property that its monopole vanishes identically. Calculating  in an arbitrary frame, using the antisymmetry of $\sigma$, and assuming  $j^\text{spn}\,{}^i(x) \rightarrow 0$ for $|\boldsymbol{x}|\rightarrow \infty$
\begin{align}
\int\d^3 \boldsymbol{x}\,j^\text{spn}{}^0 = \lim_{\epsilon\rightarrow 0} \partial_\epsilon\epsilon^{-1}\int\d^3 \boldsymbol{x}\,\int\d s\, \partial_0(\phi,\sigma^{0 0}\phi)-  \partial_i(\phi,\sigma^{i 0}\phi)=0-0=0\,  . 
\end{align}

As $\sum_k \lsuper{k}j$ generates $A$,  we clearly  have 
\begin {equation}\label{dA1}
 \partial_\nu\Theta^{\nu
 \mu} + \sum_k\, F^{\mu}_{\phantom{1}\nu}\left( \lsuper{k}j^\text{orb}\,{}^\nu+ \lsuper{k}j^\text{spn}\,{}^\nu\right)=0    \,,\quad\text{for }x\notin\bar{\gamma}\, .
\end {equation}
Repeating the procedure from section \ref{Energy-momentum conservation}  with the modified action, \eqref{A_s}, we get a conserved e-m tensor, $p$. However, the simple symmetrization trick used for the scalar case doesn't work in the current case and a symmetric, gauge invariant $p$, is more easily derived via a flat-space limit of a the fully generally covariant $p$, guaranteed to be both. This is a straightforward exercise in GR involving spinor fields.

\bibliographystyle{plain}


\end{document}